%% file: main.tex
\PassOptionsToPackage{sort,compress}{natbib}
\documentclass[final,5p,times,twocolumn,10pt]{elsarticle}

\usepackage{amsmath,amsfonts}
\usepackage{algorithmicx}
\usepackage{algorithm}
\usepackage{array}
\usepackage[caption=false,font=normalsize,labelfont=sf,textfont=sf]{subfig}
\usepackage{textcomp}
\usepackage{stfloats}
\usepackage{url}
\usepackage{verbatim}
\usepackage{siunitx}
\usepackage{cleveref}
\usepackage{xcolor}
\usepackage[acronym]{glossaries}
\usepackage{graphicx}
\usepackage{amssymb}
\usepackage{array}
\usepackage{booktabs}
\usepackage{algpseudocode}
\usepackage{setspace}

\input{settingsFigures}

\input{acronyms}

\input{def}

\journal{European Journal of Control}

\begin{document}
\begin{frontmatter}

\title{Game Theory in Formula 1: From Physical to Strategic Interactions}

\author[eth]{Giona Fieni\corref{cor1}}
\ead{gfieni@idsc.mavt.ethz.ch}

\author[eth]{Marc-Philippe Neumann}

\author[unibo]{Francesca Furia}

\author[eth]{Alessandro Caucino}

\author[ferrari]{Alberto Cerofolini}

\author[unibo]{Vittorio Ravaglioli}

\author[eth]{Christopher H. Onder}

\cortext[cor1]{Corresponding author.}
                  
\affiliation[eth]{organization={Institute for Dynamic Systems and Control, ETH Z\"urich},%
            city={8092 Z\"urich},
            country={Switzerland}}
            
\affiliation[unibo]{organization={Department of Industrial Engineering, Università di Bologna},%
            city={47121 Forlì},
            country={Italy}}
            
\affiliation[ferrari]{organization={Power Unit Performance and Control Strategies Group, Ferrari S.p.A.},%
            city={41053 Maranello},
            country={Italy}}
            
\begin{abstract}
This paper presents an optimization framework to model multi-agent racing dynamics. By incorporating physically accurate interaction models and accounting for the optimal responses of competing agents, our approach reveals strategic behaviors typical of motorsport. Aerodynamic wake effects, trajectory optimization, and energy management are captured and evaluated on a representative case study, based on a Formula 1 scenario. We describe the minimum lap time problem with two agents as either a Nash or a Stackelberg game, and by employing the Karush-Kuhn-Tucker conditions during the problem formulation, we recover the structure of a nonlinear program. In addition, we introduce an algorithm to refine local Stackelberg solutions, using the Nash costs as upper bounds. The resulting strategies are analyzed through case studies. We examine the impact of slipstreaming on trajectory selection in corners, straights, and high-speed sections, while also identifying optimal overtaking locations based on energy allocation strategies. Exploiting the structural similarities of the game formulations, we are able to compare symmetric and hierarchical strategies to analyze competitive racing dynamics. The proposed methodology closes the gap between theoretical game theory and practical applications, with relevance in multi-agent systems with coupled nonlinear dynamics.
\end{abstract}

\begin{graphicalabstract}
\includegraphics[width=15cm]{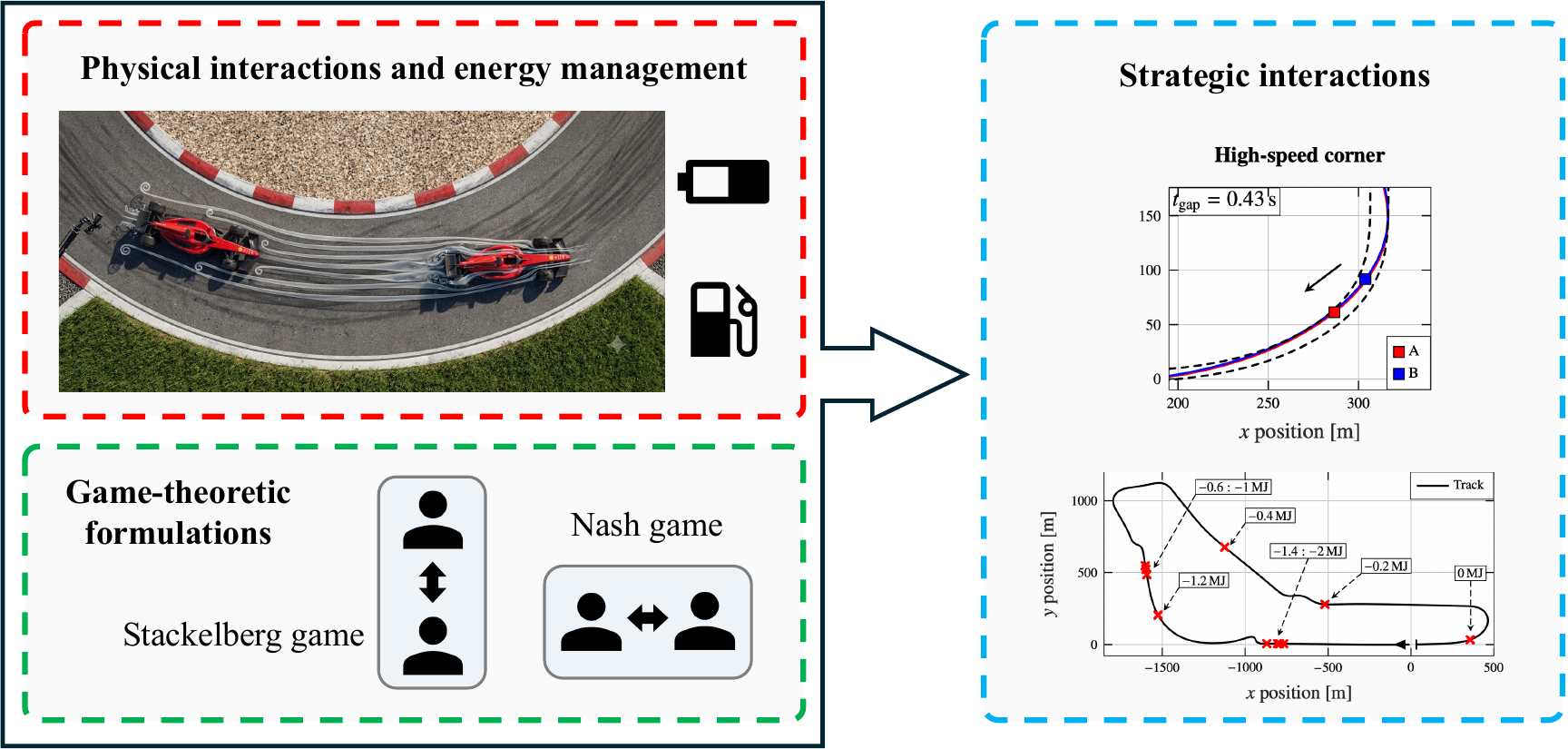}
\end{graphicalabstract}

\begin{keyword}
Energy management \sep Formula 1 \sep hybrid-electric \sep multi-agent \sep physical interactions \sep game theory \sep nonlinear programming.
\end{keyword}

\end{frontmatter}

\input{chapters/S1_Introduction.tex}

\input{chapters/S2_Modeling.tex}
\input{chapters/S3_Methodology.tex}
\input{chapters/S4_Results.tex}

\input{chapters/S5_Conclusions.tex}

\appendix
\input{chapters/S6_Appendix_1.tex}

\section*{Acknowledgments}
We thank Ferrari S.p.A. for supporting this project. Moreover, we would like to express our deep gratitude to Ilse New for her helpful and valuable comments during the proofreading phase.

\bibliographystyle{elsarticle-num} 
\bibliography{bibliography.bib}

\end{document}

%% file: settingsFigures.tex
\usepackage{tikz}
\usepackage{pgfplots}
\usepackage{graphicx}
\usepackage{standalone} %

\pgfplotsset{compat=1.18}
\usepgfplotslibrary{units}
\usetikzlibrary{positioning,intersections, hobby, patterns, calc, decorations.pathmorphing, decorations.markings, shadows, shapes, plotmarks, shapes.multipart}
\usetikzlibrary{fit}
\usetikzlibrary{arrows.meta}

\usepgfplotslibrary{external}

\tikzset{
  external/system call={pdflatex \tikzexternalcheckshellescape --halt-on-error --interaction=batchmode --jobname "\image" "\texsource"}
}
\tikzexternaldisable %

\newenvironment{externalize}[1]{
\tikzexternalenable %
\tikzsetnextfilename{img-#1} %
}{}

\pgfplotsset{
            /pgfplots/layers/niceLayers/.define layer set={
                        axis background,
                        axis grid,
                        main,
                        axis ticks,
                        axis lines,
                        axis tick labels,
                        axis descriptions,
                        axis foreground
            }{/pgfplots/layers/standard}
}

\pgfplotsset{
            every axis/.append style={
                        set layers=niceLayers,
                        tick label style={font=\scriptsize},
                        clip marker paths=true,
                        line width=1pt,
                        line cap=round,
                        line join=round,
                        tick style={semithick, color=black},
                        legend style={
                        		   font=\scriptsize,
                                    /tikz/every even column/.append style={column sep=2mm},
                                    cells={anchor=west}, %
                        },
                        xmajorgrids,
                        ymajorgrids,
            }
}

\pgfkeys{/pgf/number format/.cd,1000 sep={}}

%% file: acronyms.tex
\newacronym[plural=PUs, firstplural=power units]{pu}{PU}{power unit}
\newacronym{f1}{F1}{Formula 1}
\newacronym{cfd}{CFD}{computational fluid dynamics}
\newacronym{mpc}{MPC}{model predictive control}
\newacronym{drs}{DRS}{Drag Reduction System}
\newacronym{mguk}{MGU-K}{motor-generator unit -- kinetic}
\newacronym{mguh}{MGU-H}{motor-generator unit -- heat}
\newacronym{nn}{NN}{neural network}
\newacronym[plural=OCPs, firstplural=optimal control problems]{ocp}{OCP}{optimal control problem}
\newacronym[plural=NLPs, firstplural=nonlinear programs]{nlp}{NLP}{nonlinear program}
\newacronym{kkt}{KKT}{Karush-Kuhn-Tucker}
\newacronym{mpcc}{MPCC}{mathematical program with complementarity constraints}
\newacronym{licq}{LICQ}{linear independence constraint qualification}
\newacronym{mfcq}{MFCQ}{Mangasarian-Fromovitz constraints qualification}
\newacronym{svo}{SVO}{Social Value Orientation}
\newacronym{ibr}{IBR}{iterated best-response}
\newacronym{mltp}{MLTP}{minimum lap time problem}
\newacronym{qss}{QSS}{quasi-steady-state}
\newacronym{em}{EM}{energy management}
\newacronym[plural=GNEPs, firstplural=Generalized Nash Equilibrium Problems]{gnep}{GNEP}{Generalized Nash Equilibrium Problem}
\newacronym[plural=HEVs, firstplural=hybrid-electric vehicles]{hev}{HEV}{hybrid-electric vehicle}
\newacronym{soc}{SOC}{state-of-charge}
\newacronym{ice}{ICE}{internal combustion engine}
\newacronym{ers}{ERS}{engine recovery system}
\newacronym{rl}{RL}{reinforcement learning}

%% file: def.tex
\newtheorem{Def}{Definition}

\newtheorem{problem}{Problem}

\newcommand{\ddt}{\frac{\mathrm{d}}{\mathrm{d}t}}
\newcommand{\dds}{\frac{\mathrm{d}}{\mathrm{d}s}}
\newcommand{\isMainDocument}{}

\newlength{\myfigskip}

%% file: chapters/S1_Introduction.tex
\section{Introduction}\label{sec:intro}
Competitive motorsport racing drives innovation. Every year, the teams strive to improve and update their cars to achieve maximum performance. From aerodynamics to vehicle dynamics and \glspl{pu}, the limits of engineering are continuously pushed. 

Despite the technical innovations, the human component is still a central pillar of motorsport. Pilots exploit years of training and experience to perfectly master the vehicle, for instance by choosing the right trajectory leveraging the vehicle's dynamics. They have to take decisions in a fraction of a second by considering the presence, the actions and the response of competitors. 

Most research to date has been focused on the optimization of a single vehicle. However, the presence of other cars on the track introduces complex interactions of a both physical and strategic nature. In particular, the aerodynamic response of a vehicle is significantly influenced by the turbulence generated by a leading car. This wake effect reduces aerodynamic drag, enabling energy savings and allowing the trailing vehicle to achieve higher velocity peaks. On the other hand, the reduction in drag comes with a decrease in downforce, which can negatively impact cornering performance. While the former effect provides a competitive advantage, particularly on straights, the latter poses a challenge in high-speed corners. These trade-offs affect strategical decisions, such as overtaking maneuvers. 

With the introduction of alternative \glspl{pu} architectures, energy management has become a crucial point in winning a race. Hybrid-electric and fully electric categories directly control the performance, ensuring that the cars maintain optimal speed throughout the race without running out of energy prematurely.
 
We consider \gls{f1} as a suitable benchmark to demonstrate the proposed methodology. It represents a general case of energy management problems, which are also present, in simplified form, in other racing series. Since 2014, \gls{f1} is powered by a hybrid-electric \gls{pu}, featuring a battery, two electric motors, and a turbocharged \qty{1.6}{\liter} V6 engine. The battery has a limited capacity, refueling is forbidden, and the sporting and technical regulations \cite{2025F1, 2025F1_sport} impose further constraints. Due to the complexity of its energy management problem, optimization routines are a game changer in this context, helping the teams to develop new strategies.

In this study, we consider a multi-agent environment, featuring aerodynamic interaction, trajectory optimization, and EM optimization. Usually, these aspects are studied and treated separately. However, understanding and quantifying the interdependencies enables to further maximize the performance.

\subsection{Related work}\label{ssec:relWork}
\Cref{fig:VennDiagram} introduces the topics of interest for this work. In this section, we will address each of them, with an additional discussion of the gray areas, representing the literature gaps. Energy management is the underlying context, and its consideration in our work provides an additional value.

\textit{Trajectory optimization.}
In motorsports, trajectory optimization traditionally aims at solving the \gls{mltp}, using detailed vehicle dynamics \cite{rucco2015min,cossalter1999general,perantoni2014optimal,veneri2020free,christ2021time} for cars and motorcycles. Simplified models, such as the bicycle model, are used when computational efficiency is desired, for instance in \gls{mpc} applications \cite{kloeser2020}. On the other hand, \cite{lovato2022three,limebeer2015optimal,rowold2023optimal} consider more complex models with 3D trajectories or dynamics. A complete guide focusing on single-vehicle dynamics can be found in \cite{milliken1995race}.

Crossovers with \gls{em} or aerodynamic interactions are studied in \cite{tian2024coordinated} and \cite{liu2024optimal}, respectively. The first considers drones, but the problem is divided into two layers:  offline trajectory optimization and online \gls{em}, which maximizes the energy produced by photovoltaic cells. The second optimizes the trajectory of a race car by considering the slipstream effect, although the car in front is fixed and does not interact with the other one. Additionally, only the longitudinal reduction of drag and downforce is taken into account.
 
\textit{Aerodynamic interaction.}
Aerodynamic wake effects in \gls{f1} cars have been widely studied in static \gls{cfd} simulations \cite{ravelli2021aerodynamics,mafi2007investigation,dvzijan2021aerodynamic,guerrero2020aerodynamic} and experimental analyses \cite{dominy1990influence,soso2006aerodynamics,newbon2017aerodynamic}. Whilst the drag and downforce losses are often investigated for different vehicle longitudinal spacing, less effort is spent to describe the reduction given by the lateral spacing \cite{dominy1990influence,ravelli2021aerodynamics,gan2020cfd}. 

The work in \cite{fieni2024game} represents the first step towards the inclusion of aerodynamic interaction effects into multi-agent dynamical systems. It integrates a longitudinal drag reduction model, inspired by \gls{cfd} literature, and assessed its influence on the \gls{em} of two \gls{f1} cars within a \gls{mltp} formulation. However, it does not account for lateral drag reduction and downforce loss, important factors impacting the optimal trajectory choice. 

Subsequently, longitudinal and lateral drag interaction were explored in \cite{limebeer2025optimal}, to study their impact on an overtaking maneuver. While this later work incorporates a broader aerodynamic model, it performs a posteriori an optimization of the overtaking vehicle given a fixed trajectory for the lead car. However, the agent ahead has no awareness of the overtaking agent and cannot react.

\textit{Multi-agent systems.}
Although there is an extensive literature on multi-agent dynamical systems, we distinguish between game-theoretic methods and \gls{rl}. While game theory literature is oriented at finding optimal pure or mixed strategies, the \gls{rl} literature focuses on stochastic policies and robustness in motion planning. In this work, we leverage game theory to find optimal pure strategies, which are particularly suited for energy management applications, where precise control sequences are required.

Game-theoretic approaches in robotics traditionally use receding horizon best-response algorithms for drones and cars autonomous racing \cite{williams2018best,wang2019game,wang2019game2,wang2020multi,wang2021game,notomista2020enhancing,spica2020real,liniger2019noncooperative}. The goal is to sequentially solve the path planning problem while avoiding collisions with other agents. Regarding social awareness, \cite{schwarting2019social,burger2022interaction} highlight cooperative interactions, whereas \cite{spica2020real,liniger2019noncooperative} apply noncooperative games to autonomous competitive racing. 

A particular focus on trajectory optimization in game-theoretic frameworks is provided in \cite{cui2023beat,le2022algames}. The first integrates a tire model for the vehicle motion within an iterative best-response algorithm. The second employs a root-finding algorithm with an augmented Lagrangian method to handle trajectory optimization in a multi-agent environment.

The intersection with aerodynamic interaction is studied in \cite{fieni2024game,cinar2024does}. The first integrates a model based on literature data, while the second uses a rudimentary drag interaction model. Both address hierarchical leader-follower dynamics inherent to competitive racing. However, \cite{fieni2024game} optimizes an entire \gls{f1} lap, whereas \cite{cinar2024does} relies on a receding horizon approach and simulates only a few steps. Although none of them directly compares Nash with Stackelberg equilibria, \cite{cinar2024does} analyzes the costs resulting from the generated closed-loop strategies.
 
\textit{Energy management in racing.}
\gls{em} in hybrid-electric racing cars has been widely studied. Both offline optimization \cite{ebbesen2017time,duhr2022convex,duhr2023minimum,balerna2020optimal,balerna2020time} and online control \cite{salazar2017time,salazar2017real,neumann2023low} have been investigated. In particular, \cite{balerna2020optimal} focuses on optimal fuel-efficiency operation, \cite{balerna2020time} on time-optimal operation, \cite{duhr2023minimum} on time-optimal gearshift strategies. In \cite{duhr2022convex} the authors include a g-g diagram to study the effect on the \gls{em} for a predefined trajectory. Energy-optimal overtaking maneuvers are addressed in \cite{liu2023energy} for Formula E cars. However, this research lacks a description of active interaction between competitors. Further overlaps of \gls{em} with the other fields have already been discussed in this section. 

\subsection{Problem statement}
The objective of this paper is to close the gaps in the current race car optimization methodologies. Typically, the \gls{em}, a multi-agent setting, the choice of the racing line and the inclusion of aerodynamic interactions are analyzed separately in the literature. The limited capability of existing optimization tools fails to capture the complex, interdependent, and competitive nature of racing scenarios. 

Extending the work \cite{fieni2024game}, we propose a holistic approach to optimize the interplay of all these features. We describe two agents racing on the same track with their mutual aerodynamic influence, including optimal reactions to other agents' strategies according to the game formulation. Computing open-loop game equilibria, we investigate how strategic plans can emerge from the physical coupling.

We provide a robust and efficient approach that will enable teams to achieve higher performance by optimizing all the factors together. 

\subsection{Contributions}\label{contributions}
\begin{figure}
\centering
\scalebox{0.8}{
\begin{externalize}{VennDiagram}
\input{figures/VennDiagram/VennDiagramV2.tex}
\end{externalize}
}
\caption{Representation of the topics covered in this paper. The literature gaps that we aim to address are highlighted in gray. The numbering corresponds to the contributions outlined in \Cref{contributions}.}
\label{fig:VennDiagram}
\end{figure}

To tackle these challenges, we develop a novel framework in this paper, to optimize multi-agent racing strategies. In particular, we contribute in three distinct ways, highlighted in \Cref{fig:VennDiagram}. The literature gaps that we address are marked in gray. 

First, we integrate \gls{em}, aerodynamic interactions models, and trajectory optimization into a multi-agent dynamic framework. Our approach dynamically couples wake-induced aerodynamics changes, collision avoidance, and trajectory with hybrid powertrain energy allocation, enabling accurate lap time optimizations. 

Second, we include these models in a optimization-based game-theoretic framework. By capturing symmetric competition or hierarchical leader-follower dynamics inherent to motorsport, we derive optimal racing strategies directly emerging from the game formulation. The similar problem structures allow for a direct comparison of the different games' outcomes across different scenarios. Additionally, we propose a method to refine local Stackelberg solutions by leveraging a property of game theory. 

Third, we showcase the impact of the holistic approach on the strategical behavior stemming from the interactions. For instance, we show the balance between trajectory choice and wake effect or between optimal overtake location and energy target.

We validate these contributions by means of case studies, in order to assess strategies of real-world racing scenarios. Eventually, these contributions bridge theoretical and practical gaps, offering a tool to optimize racing strategies under complex multi-agent interactions.

\subsection{Outline}
This paper is structured as follows: In \cref{sec:modeling}, we introduce the single-agent dynamic models, along with drag and downforce reduction models, collision avoidance constraints, and the trajectory model. Then, we formulate the \gls{ocp} and the resulting \gls{nlp}. In \cref{sec:method}, we describe in detail the different game-theoretic approaches that can be applied to the problem formulation. The results are presented and discussed in \cref{sec:results}. Finally, we draw conclusions in \cref{sec:conclusion}, highlighting the relevant insights of our work and present an outlook on future research.

%% file: figures/VennDiagram/VennDiagramV2.tex
\begin{tikzpicture}%
\tikzset{>=stealth}

\def\radius{1.7cm}
\def\offset{1.5*\radius}
\def\offsetText{0.33*\radius}
\def\secondGroupOffset{2.5*\radius}

\node at (0,0) (leftCircle) {};
\node at ($(leftCircle) + (0.5*\offset, 0.866*\offset)$) (upperCircle) {};
\node at (\offset, 0) (rightCircle) {};
\node at ($(leftCircle) + (0.75*\radius, -1.03*\radius)$) (midpoint) {};

\node at ($(0,0) + (0, -\secondGroupOffset)$) (leftCircle2) {};
\node at ($(leftCircle2) + (\offset, 0)$) (rightCircle2) {};
\node at ($(leftCircle2) + (0.75*\radius, 0)$) (midpoint2) {};

\begin{scope}
 \clip (0,0) circle (\radius);
 \fill[gray!70] (upperCircle) circle (\radius);
 \fill[gray!70] (rightCircle) circle (\radius);
\end{scope}

\begin{scope}
 \clip (upperCircle) circle (\radius);
 \fill[gray!70] (rightCircle) circle (\radius);
\end{scope}

\begin{scope}
 \clip (rightCircle2) circle (\radius);
 \fill[gray!70] (leftCircle2) circle (\radius);
\end{scope}

\draw (0,0) circle (\radius);
\draw (rightCircle) circle (\radius);
\draw (upperCircle) circle (\radius);

\draw (leftCircle2) circle (\radius);
\draw (rightCircle2) circle (\radius);

\draw[draw=black, line width=1.5pt] ($(leftCircle) + (-1.1*\radius, 2.4*\radius)$) rectangle ($(rightCircle) + (1.1*\radius, -1.1*\radius)$);

\draw[draw=black, line width=1.5pt] ($(leftCircle2) + (-0.6*\radius, -1.6*\radius)$) rectangle node[midway] {Strategic interactions} ($(rightCircle2) + (0.6*\radius, -2.2*\radius)$);

\draw[->, draw=black, line width=1.5pt] (midpoint) -- (midpoint2);

\draw[->, draw=black, line width=1.5pt] ($(midpoint2) + (0, -0.78*\radius)$) -- ($(midpoint2) + (0, -1.55*\radius)$);

\node[align=center] at (upperCircle) {Energy\\management};
\node[align=center] at ($(rightCircle) + 0.5*(\offsetText, 0)$) {Trajectory\\optimization};
\node[align=center] at ($(0,0) + 0.5*(-\offsetText, 0)$) {Aerodynamic\\interactions};

\node[align=center] at ($(rightCircle2) + (0.6*\offsetText, 0)$) {Game theory\\(multi-agent)};
\node[align=center] at ($(leftCircle2) + (-0.6*\offsetText, 0)$) {Optimization};

\node[align=center, draw, circle] at ($(leftCircle) + (-0.8*\radius, 2.1*\radius)$) {1.};
\node[align=center, draw, circle] at ($(midpoint) + (-0.17*\offset, -0.22*\offset)$) {2.};
\node[align=center, draw, circle] at ($(midpoint2) + (-0.17*\offset, -0.8*\offset)$) {3.};

\end{tikzpicture}

%% file: chapters/S2_Modeling.tex
\section{Modeling}\label{sec:modeling}

In this section, we present a model to perform multi-agent optimizations on a race track. We first introduce single-agent dynamic models which account for individual vehicle dynamics, including \gls{pu} model and trajectory optimization. Subsequently, we extend this model to incorporate interactions with a competitor. In particular, we focus on aerodynamic coupling, where slipstream and turbulent wake effects generated by one agent directly influence the other. Furthermore, we integrate collision avoidance constraints into the formulation, replicating drivers actively avoiding collisions during competition. The combined framework enables the study of interactions between agents simultaneously optimizing their performance while dynamically responding to aerodynamic disturbances and spatial conflicts.
 
\begin{figure}
	\centering
	\includegraphics[width=0.75\columnwidth]{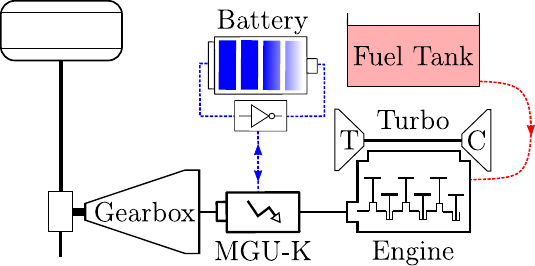}
	\caption{Schematic of the \gls{f1} \gls{pu}. The on-board energy storages are the fuel tank and the battery. The prime movers are the \gls{mguk} and the turbocharged engine.}
	\label{fig:F1PU}
\end{figure}

Prior to detailing the model, we briefly describe the \gls{f1} hybrid-electric \gls{pu}. It is a system which combines a \qty{1.6}{\liter} turbocharged V6 \gls{ice} with an \gls{ers}. Their respective energy storages are a fuel tank and a battery. The \gls{ers} is composed of two electric motors, the \gls{mguk} and the \gls{mguh}. In particular, the \gls{mguk} recovers kinetic energy during braking, converting it into electrical energy that can be stored in the battery and later used to provide additional power to the drivetrain. The \gls{mguh}, on the other hand, is connected to the turbine and recovers thermal energy from the exhaust gases. By converting it into electrical energy, it can be stored and used to reduce turbo lag by maintaining the turbocharger's speed. This hybrid configuration allows for a more efficient use of fuel and energy, enhancing the overall performance and sustainability of the race car. To reduce mathematical complexity, our model neglects the \gls{mguh} as shown in \Cref{fig:F1PU}. 

\subsection{Single-agent model}
We now describe the single-agent dynamic model. Similarly to \cite{ebbesen2017time}, the problem is formulated in space domain, where we use the centerline curvilinear coordinate $s\in\{0,S\}$ as an independent variable, with $S$ representing the track length. The track is characterized by its curvature $\gamma(s)$ and slope $\theta(s)$. This space domain formulation is adopted because the track characteristics are inherently spatial, and because discretization in space domain ensures a fixed number of optimization steps. Conversely, time-domain discretization would yield variable steps due to velocity-dependent grid resolution.

In this work, the car is modeled as a point mass, neglecting yaw and out-of-plane dynamics. While factors such as banking angles and vertical dynamics can influence local trajectory adjustments, they are omitted here. This reduced-order approach maintains computational efficiency while capturing the dominant physical effects, for instance the impact that aerodynamic interactions have on the \gls{em}.

The state vector \(\mathbf{x}\) includes the vehicle's velocity \(v\), fuel energy \(E_\mathrm{f}\), the battery energy \(E_\mathrm{b}\), elapsed time \(t\), energy recovered by the \gls{mguk} and stored in the battery \(E_\mathrm{K2ES}\), lateral displacement from the track centerline \(y\), and heading angle \(\varphi\), which represents the vehicle's orientation with respect to the centerline tangent 
\begin{equation}
	\mathbf{x} = \begin{bmatrix}
		v & E_\mathrm{f} & E_\mathrm{b} & t & E_\mathrm{K2ES} & y & \varphi
	\end{bmatrix}^\intercal.
\end{equation}
The control inputs \(\mathbf{u}\) include fuel power \(P_\mathrm{f}\), \gls{mguk} mechanical power \(P_\mathrm{k}\), braking power \(P_\mathrm{brk}\), electrical power recovered by the \gls{mguk} \(P_\mathrm{K2ES}\), and lateral acceleration \(a_\mathrm{lat}\). They are subject to regulatory limits, except for braking power and lateral acceleration. The control vector is
\begin{equation} 
	\mathbf{u} = \begin{bmatrix}
		P_\mathrm{f} & P_\mathrm{k} & P_\mathrm{brk} & P_\mathrm{K2ES} & a_\mathrm{lat}
	\end{bmatrix}^\intercal.
\end{equation}

The system dynamics are governed by the following ordinary differential equations: 
\begin{equation}\label{eq:dyneq}
	\ddt \mathbf{x}(t) = 
	\begin{cases}
		\ddt v(t)  &=  \frac{1}{m} \cdot \frac{ P_\mathrm{p}(t) - P_\mathrm{ext}(t)}{v(t)} \\
		\ddt E_\mathrm{f}(t)   &=  P_\mathrm{f}(t) \\
		\ddt E_\mathrm{b}(t)   &=  -P_\mathrm{i}(t) \\
		\ddt t   &=  1, \\\ddt E_\mathrm{K2ES}(t)  &=  -P_\mathrm{K2ES}(t) \\
		\ddt y(t)   &=  v(t) \cdot \sin(\varphi(t)) \\		
		\ddt \varphi(t)  &=  \frac{a_\mathrm{lat}(t)}{v(t)} - \ddt s(t) \cdot \gamma(s)
	\end{cases},
\end{equation}
where \(m\) is the vehicle mass, \(P_\mathrm{p}\) is the net propulsive power, \(P_\mathrm{ext}\) aggregates external power losses, and  \(P_\mathrm{i}\) is the internal battery power. To switch from time to space domain, we use the transformation
\begin{equation}\label{eq:dds}
	\frac{\mathrm{d}s}{\mathrm{d}t}= v_\mathrm{c}(t) \quad  \Rightarrow \quad \mathrm{d}t = \frac{\mathrm{d}s}{v_\mathrm{c}(s)},
\end{equation}
where $v_\mathrm{c}$ is the velocity projected along the centerline. 
According to \cite{kloeser2020}, the velocity along the track's centerline is 
\begin{equation}
	v_\mathrm{c}(t) = \frac{v(t) \cdot \cos(\varphi(t))}{1-y(t) \cdot \gamma(s)}.
\end{equation}
The set of dynamic equations is then converted as
\begin{equation}\label{eq:spaceconv}
	\ddt \mathbf{x}(t)= F(t) \quad \Rightarrow \quad \dds \mathbf{x}(s)= \frac{F(s)}{v_\mathrm{c}(s)},
\end{equation}
where $F(\cdot)$ is the right-hand side of \cref{eq:dyneq}.

\subsubsection{Boundary conditions}
First, we introduce the model boundary conditions. We consider the battery energy at the beginning of the lap, denoted as \(E_\mathrm{b,0}\), as a predefined parameter. The variation in energy within the lap is defined by a target \(\Delta{E_\mathrm{b,target}}\):   
\begin{alignat}{2}\label{eq:eb}
	E_\mathrm{b}(0) &= E_\mathrm{b,0}, \nonumber\\
	E_\mathrm{b}(S) &\ge E_\mathrm{b}(0) + \Delta{E_\mathrm{b,target}}. 
\end{alignat}
Regulations limit the amount of energy per lap that can be recovered by the battery from the \gls{mguk}: 
\begin{alignat}{2}\label{eq:ek2es}
	E_\mathrm{K2ES}(0) &= 0, \nonumber\\
	E_\mathrm{K2ES}(S) &\leq E_\mathrm{K2ES,max}. 
\end{alignat}
This energy budget is initialized to zero to account for the reset occurring at the beginning of each lap.
The fuel energy allocated per lap is strategically limited to ensure enough fuel for the entire race:
\begin{alignat}{2}\label{eq:efmax}
	E_\mathrm{f}(S) &\leq E_\mathrm{f,max}. 
\end{alignat}
Similarly, this energy budget is reset at the start of each lap: 
\begin{alignat}{2}\label{eq:ef}
	E_\mathrm{f}(0) &= 0.
\end{alignat}
While the total laptime is subject to optimization, the initial time is a boundary condition: 
\begin{alignat}{2}\label{eq:tinit}
 	t(0) &= 	t_\mathrm{init}.
\end{alignat}
This initialization is particularly important when analyzing multi-agent interactions, as it defines the initial gap between the vehicles. 

\subsubsection{Power Unit model} 
Net propulsive power \(P_\mathrm{p}\) is the power contributing to vehicle motion. It combines the power coming from the gearbox \(P_\mathrm{g}\) with braking power:
\begin{equation}\label{eq:pp}
	P_\mathrm{p}(s) = P_\mathrm{g}(s) - P_{\mathrm{brk}}(s).
\end{equation}
We account for gearbox inefficiency with
\begin{equation} \label{eq:pg}
	P_\mathrm{g}(s) = a_\mathrm{g} \cdot P_\mathrm{u} ^2(s) + P_\mathrm{u}(s),
\end{equation} 
where \(P_\mathrm{u}\) is the \gls{pu} power and \(a_\mathrm{g}<0\). \(P_\mathrm{u}\) combines the internal combustion engine power \(P_\mathrm{e} \) and the mechanical \gls{mguk} power \(P_\mathrm{k} \):
\begin{equation} \label{eq:pu}
	P_\mathrm{u}(s) = P_\mathrm{e}(s) + P_\mathrm{k}(s).
\end{equation} 
The engine power \( P_\mathrm{e}\) is modeled as 
\begin{equation} \label{eq:pe}
	P_\mathrm{e} (s) = \eta_e \cdot P_\mathrm{f} (s) -  P_\mathrm{e,0},
\end{equation} 
where \(\eta_e\) is the Willans efficiency, assumed constant, and \( P_\mathrm{e,0}\) accounts for frictional and pumping losses \cite{guzzella2009introduction}. The \gls{mguk} electrical-to-mechanical and mechanical-to-electrical inefficiency is taken into account with the quadratic equation 
\begin{equation} \label{eq:pkdc}
	P_\mathrm{k,dc}(s)  = a_\mathrm{k} \cdot P_\mathrm{k} ^2(s) + P_\mathrm{k}(s), 
\end{equation}
where \(P_\mathrm{k,dc} \) is the \gls{mguk} electrical power and $a_\mathrm{k}>0$. 
The system is subject to the following constraints: 
\begin{alignat}{4}\label{eq:constr}
	0 &\quad \leq \quad P_\mathrm{f}(s) &\quad \leq \quad &P_\mathrm{f,max}, \nonumber\\
	P_\mathrm{k,dc,min} &\quad \leq \quad P_\mathrm{k,dc}(s) &\quad \leq \quad &P_\mathrm{k,dc,max},\nonumber\\
	0 &\quad \leq \quad P_\mathrm{brk}(s) &\quad \leq \quad &P_\mathrm{brk,max},\nonumber\\
	P_\mathrm{k,dc,min} &\quad \leq \quad P_\mathrm{K2ES}(s) &\quad \leq \quad &0,\nonumber\\
	0 &\quad \leq \quad E_\mathrm{b}(s) &\quad \leq \quad &E_\mathrm{b,max},\nonumber\\
	0 &\quad \leq \quad E_\mathrm{K2ES}(s) &\quad \leq \quad &E_\mathrm{K2ES,max},
\end{alignat}
The battery power \( P_\mathrm{b}\) includes auxiliary loads \( P_\mathrm{aux}\):
\begin{equation} \label{eq:pb}
	P_\mathrm{b}(s) =  P_\mathrm{k,dc}(s) +  P_\mathrm{aux}.
\end{equation} 
The internal battery power \( P_\mathrm{i}\) is modeled as
\begin{equation}\label{eq:pi}
	P_\mathrm{i}(s)  = a_\mathrm{b} \cdot  P_\mathrm{b} ^2(s) +  P_\mathrm{b}(s), 
\end{equation} 
where \(a_\mathrm{b}> 0\) accounts for battery charge/discharge losses.

External power \(P_\mathrm{ext}\) includes aerodynamic drag \(P_\mathrm{aero}\), rolling resistance  \(P_\mathrm{roll}\) and slope effects \(P_\mathrm{slope}\): 
\begin{equation}\label{eq:pext}
	P_\mathrm{ext}(s) = P_\mathrm{aero}(s) +  P_\mathrm{roll}(s) +  P_\mathrm{slope}(s). 
\end{equation} 
The aerodynamic drag power is
\begin{equation}\label{eq:paero}
	P_\mathrm{aero}(s)  =  c_\mathrm{d,1}\cdot v^3(s) + c_\mathrm{d,2} \cdot a_\mathrm{lat}(s)\cdot v(s),
\end{equation} 
and is expressed as a function of the drag coefficient $c_\mathrm{d,1}$ and a curvature-dependent term $c_\mathrm{d,2}$, for which we used the curvature of the current racing line $a_\mathrm{lat}(s)/v(s)^{2}$. The latter term models sidewind effects due to the open-wheel design of the vehicle \cite{ebbesen2017time}. The rolling resistance is expressed as
\begin{equation}\label{eq:proll}
	P_{\mathrm{roll}}(s) = c_{\mathrm{roll}} \cdot m \cdot g \cdot \cos(\theta(s)) \cdot v(s),
\end{equation} 
with \(c_{\mathrm{roll}}\) being the rolling resistance coefficient, and \(g\) the gravitational acceleration. Finally,
\begin{equation}\label{eq:pslope}
	P_{\mathrm{slope}}(s) = m \cdot g  \cdot \sin(\theta(s)) \cdot v(s).
\end{equation} 

\subsubsection{Performance envelope} 
Longitudinal and lateral forces acting on the vehicle are computed as 
\begin{alignat}{2}
	\label{eq:flongdef} F_\mathrm{long}(s) &= \frac{P_\mathrm{p}(s)}{v(s)},  \\
	\label{eq:flatdef} F_\mathrm{lat}(s) &= m \cdot a_\mathrm{lat}(s),   
\end{alignat} 
with \(a_\mathrm{lat}(s)\) subject to 
\begin{equation} \label{eq:acclat}
	a_\mathrm{lat,min} \quad \leq  \quad a_\mathrm{lat}(s) \quad \leq \quad a_\mathrm{lat,max}.
\end{equation}
To ensure the vehicle's stability, lateral and longitudinal forces are constrained by tire grip limits derived from track-dependent parameters \cite{duhr2022convex}: 
\begin{alignat}{2}\label{eq:perfenv2}
	\left(\frac{F_\mathrm{lat}(s)}{F_\mathrm{lat,max}(s)}\right)^2 + \left(\frac{F_\mathrm{long,acc}(s)}{F_\mathrm{long,max,acc}(s)}\right)^2 &\leq 1, \\
	\label{eq:perfenv1}	\left(\frac{F_\mathrm{lat}(s)}{F_\mathrm{lat,max}(s)}\right)^2 + \left(\frac{F_\mathrm{long,dec}(s)}{F_\mathrm{long,max,dec}(s)}\right)^2 &\leq 1, 
\end{alignat}
with the maximum admissible forces defined as: 
\begin{alignat}{3} \label{eq:perfenv}
	F_{\mathrm{lat,max}}(s) &=  \alpha_{\mathrm{lat},2}\cdot v(s)^2 + \alpha_{\mathrm{lat},1}\cdot v(s) + \alpha_{\mathrm{lat},0},  \nonumber\\
	F_{\mathrm{long,max,acc}}(s) &=\beta_{\mathrm{acc},2}\cdot v(s)^2 + \beta_{\mathrm{acc},1}\cdot v(s) + \beta_{\mathrm{acc},0},  \nonumber\\
	F_{\mathrm{long,max,dec}}(s) &= \beta_{\mathrm{dec},2}\cdot v(s)^2 + \beta_{\mathrm{dec},1}\cdot v(s) + \beta_{\mathrm{dec},0},  
\end{alignat}
where \(\alpha_\mathrm{(\cdot)}\) and \(\beta_\mathrm{(\cdot)}\) are identified coefficients. The asymmetry between acceleration and deceleration limits stems from dynamic weight transfer. To link \cref{eq:perfenv2,eq:perfenv1} to \cref{eq:flongdef} we reformulate:
\begin{alignat}{4}
	\label{eq:flong} F_\mathrm{long}(s) &\quad = \quad F_\mathrm{long,acc}(s) &\quad + \quad &F_\mathrm{long,dec}(s), \\ 
	\label{eq:flongacc} 0 &\quad \leq \quad  F_\mathrm{long,acc}(s) &\quad \leq \quad &\infty,\\
	\label{eq:flongdec} -\infty  &\quad \leq \quad F_\mathrm{long,dec}(s) &\quad \leq \quad &0, 
\end{alignat}

\subsubsection{Trajectory model} 
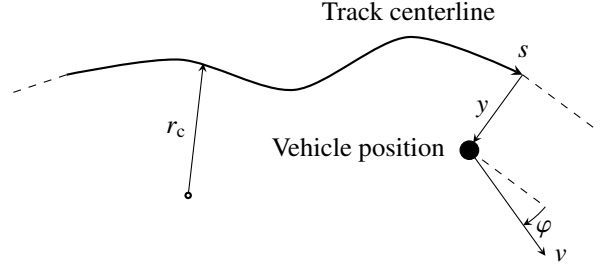
\begin{figure}
\centering
	\begin{externalize}{TrajectoryFrame}
		\input{figures/TrajectoryFrame/TrajectoryFrame.tex}
	\end{externalize}
	\caption{Schematic of the trajectory's model. The centerline curvilinear coordinate is represented by $s$, $r_\mathrm{c}$ is the curvature radius, $y$ the lateral displacement of the agent w.r.t. the centerline and $\varphi$ the heading angle.}
	\label{fig:TrajectoryFrame}
\end{figure}

The trajectory optimization problem is included in the framework by means of the last two dynamic equations in (\ref{eq:dyneq}) converted into space domain through (\ref{eq:dds}). 
\begin{alignat}{2}
	\dds y(s)   &=   \tan(\varphi(s)) \cdot \left(1-y(s) \cdot \gamma(s)\right), \\		
	\dds \varphi(s)  &=  \frac{a_\mathrm{lat}(s)}{v(s)^2} \cdot \frac{1-y(s) \cdot \gamma(s)}{\cos(\varphi(s))} - \gamma(s).
\end{alignat}
They define the rate of change of the lateral position \(y\) and the rate of change of the vehicle orientation \(\varphi\), respectively. The track's curvature is defined as: 
\begin{equation}
	\gamma(s) = 1/r_\mathrm{c}(s),
\end{equation}
where \(r_\mathrm{c}(s)\) is the curvature radius. \Cref{fig:TrajectoryFrame} shows the trajectory model with the coordinate system.

To ensure the vehicle remains on the track, the vehicle's lateral displacement is constrained by track boundaries: 
\begin{alignat}{3}\label{eq:tracklim}
	y_\mathrm{min}(s) &\quad \leq \quad y(s) &\quad \leq \quad y_\mathrm{max}(s).
\end{alignat}

We neglect the vehicle sideslip angle reducing computational complexity while maintaining fidelity. This assumption is valid in high-speed trajectories where we want to prioritize lateral grip dominance, inherently minimizing tire slip angles \cite{pacejka2012tire}. 

\subsection{Multi-agent interactions}
\begin{figure}
\centering
	\begin{externalize}{RelGap}
		\input{figures/RelGap/RelGap.tex}
	\end{externalize}
	\caption{Schematic example of the agents in the coordinate system. We show two locations along the track, with lateral deviations and time. As an example, for the time instant $t = \qty{8}{\second}$, $A$ is at $s_{2}$, ahead of $B$, which is still at $s_{1}$. The lateral deviation is computed when the cars are at the same location, e.g., $s_{1}$.}
	\label{fig:RelGap}
\end{figure}
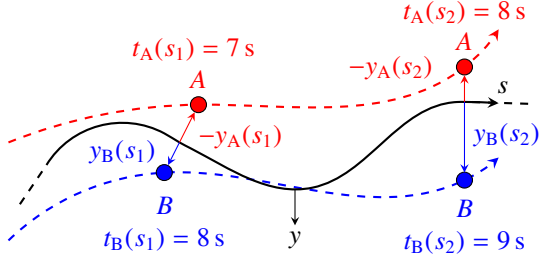

In this section, we describe how the single-agent optimal control problem is extended to include interactions between two vehicles. In particular, we include aerodynamic coupling and collision avoidance constraints. 
We consider two agents $A$ and $B$ and use index $i$ to indicate ``agent $i$'' and $-i$ to indicate ``not agent $i$''. For $i\in\{A,B\}$, the relative time gap \(t_{\mathrm{gap,rel},i}\) and relative lateral gap \(y_{\mathrm{gap,rel},i}\) are defined as
\begin{alignat}{2}
	t_{\mathrm{gap,rel},i}(s) &= t_{i}(s) - t_{-i}(s),  \\
	y_{\mathrm{gap,rel},i}(s) &= y_{i}(s) - y_{-i}(s),
\end{alignat}
where \(t_{i}\) and \(y_{i}\) denote the elapsed time and lateral displacement of agent $i$. \Cref{fig:RelGap} displays examples of $t_{i}(s)$ and $y_{i}(s)$ for two different agents for illustration purposes. For the sake of readability, the subscript $i$ is omitted in subsequent equations.

\subsubsection{Collision avoidance}
To prevent collisions, we enforce a minimum distance between the two vehicles. This separation is modeled as an elliptical constraint,
\begin{equation}\label{eq:collision }
	\left(\frac{t_\mathrm{gap,rel}(s)}{t_\mathrm{gap,min}}\right)^2 + \left(\frac{y_\mathrm{gap,rel}(s)}{y_\mathrm{gap,min}}\right)^2 \geq 1, 
\end{equation}
where \(t_\mathrm{gap,min}\) and \(y_\mathrm{gap,min}\) are the minimum allowable gaps. The lateral minimum gap corresponds to vehicle width, while \(t_\mathrm{gap,min}\) is the minimum longitudinal gap expressed in terms of time. The longitudinal gap threshold \(t_\mathrm{gap,min}=\qty{0.1}{\second}\) is chosen based on empirical studies \cite{dominy1990influence} showing peak aerodynamic drag and downforce reduction at this gap time. Despite the fixed length of the vehicles, drivers generally avoid gaps below \(\qty{0.1}{\second}\) due to excessive proximity to the other vehicle. This critical threshold aligns with real-world racing behavior, where overtaking maneuvers typically initiate above $\qty{0.1}{\second}$ to exploit maximum slipstream benefits.  

\subsubsection{Drag interactions}
\begin{figure}
\input{figures/DragReduction/DragReduction.tex}
\caption{Longitudinal (top) and lateral (bottom) drag reduction coefficients, as a function of the relative gap time and the normalized lateral gap, respectively. The points represent the data extracted from the literature, whereas the solid black lines are the fitted model.}
\label{fig:DragReduction}
\end{figure}
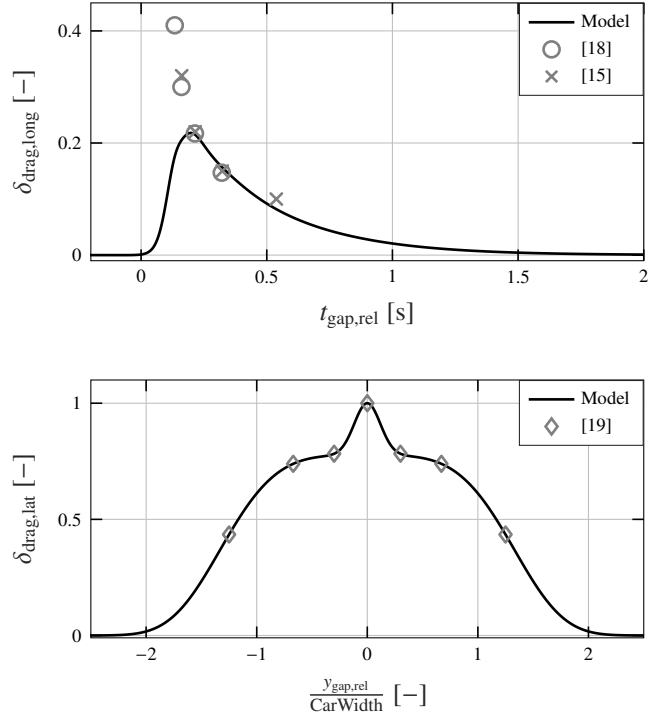

When vehicles race in close proximity, the car in front disrupts the airflow, reducing aerodynamic drag and downforce of the trailing vehicle. Following the approach in \cite{fieni2024game}, we extend the interaction model to account for downforce reduction and reduction caused by lateral misalignment. Longitudinal interactions are modeled as a function of the \textit{temporal} gap, while lateral interactions are modeled as a function of the \textit{spatial} gap. The former choice accounts for the fact that, at a given distance, the intensity of flow perturbations scales proportionally with velocity. Further details are provided in Section II-B of \cite{fieni2024game}. Our formulation assumes a point-mass vehicle model, neglecting spatial variations in the aerodynamic pressure center caused by wake effects; only the net force magnitude is scaled. $P_\mathrm{aero,int}$ captures the aerodynamic power caused by the interaction: 
\begin{alignat}{3}
	\label{eq:paeroint} P_\mathrm{aero,tot}(s)  &=  P_\mathrm{aero}(s) - 	P_\mathrm{aero,int}(s), \\ 
	P_\mathrm{aero,int}(s) &= C_\mathrm{x,int}(s) \cdot c_\mathrm{d,1} \cdot v^3(s),  \label{eq:aeroint}\\ 
	C_\mathrm{x,int}(s) &= \delta_\mathrm{drag,long}(s) \cdot \delta_\mathrm{drag,lat}(s). \label{eq:cx}
\end{alignat} 
Here, \(\delta_\mathrm{drag,long}\) represents the drag reduction factor due to the vehicle's relative gap time \(t_\mathrm{gap,rel}\), and \(\delta_\mathrm{drag,lat}\) represents the reduction due to lateral offset between their central axes \(y_\mathrm{gap,rel}\). Both factors are fitted using \gls{nn} techniques as described in \cite{balerna2020time}, employing nonlinear activation functions. This approach results in smooth and twice differentiable functions, making them suitable for the solver. The fittings are described by
\begin{alignat}{2}	\label{eq:cx1}
	\delta_\mathrm{drag,long}(s) &= \mathcal{M_\mathrm{1}}(t_\mathrm{gap,rel}(s)), \\
\label{eq:cx2}	\delta_\mathrm{drag,lat}(s) &= \mathcal{M_\mathrm{2}}(y_\mathrm{gap,rel}(s)),
\end{alignat}
where $\mathcal{M}$ denotes the \gls{nn} function. The fittings are illustrated in \Cref{fig:DragReduction}. 
According to \cite{newbon2017aerodynamic}, longitudinal and lateral aerodynamic effects are treated independently. The longitudinal factor \(\delta_\mathrm{drag,long}\) is obtained by fitting \cite{guerrero2020aerodynamic,ravelli2021aerodynamics} with a smooth function. This factor peaks at $t_\mathrm{gap,rel} =\qty{0.2}{\second}$ to ensure a smooth decrease to zero for negative gaps. The lateral drag factor \(\delta_\mathrm{drag,lat}\) is derived from  \cite{dominy1990influence} assuming symmetric vehicles, making the model symmetric around \(y_\mathrm{gap,rel}(s)=0\). 

\subsubsection{Downforce interactions}
\begin{figure}
\input{figures/DownforceReduction/DownforceReduction.tex}
\caption{Longitudinal (top) and lateral (bottom) downforce reduction coefficients, as a function of the relative gap time and the normalized lateral gap, respectively. The points represent the data extracted from the literature, whereas the solid black lines are the fitted model.}
\label{fig:DownReduction}
\end{figure}
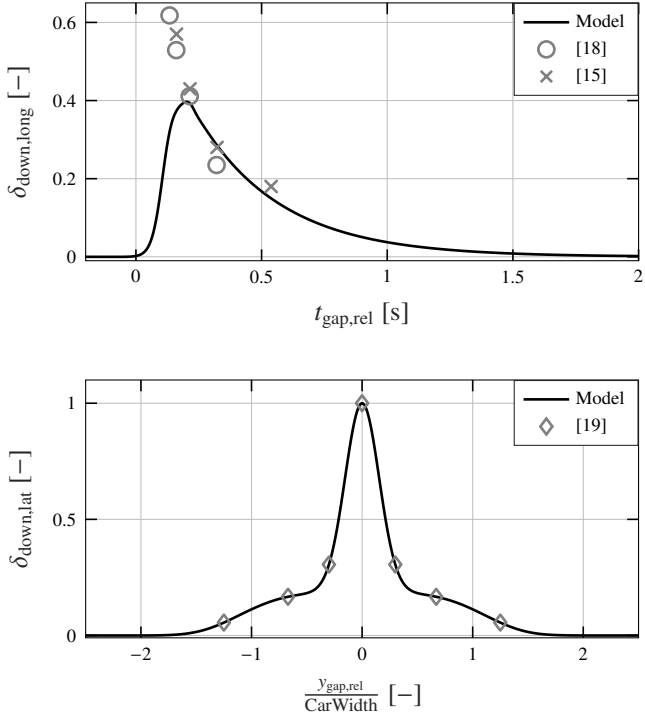

The downforce reduction coefficient $C_\mathrm{z,int}(s)$ is formulated analogously to the drag reduction, using separable longitudinal and lateral scaling factors:

\begin{alignat}{2} 	\label{eq:cz}
	C_\mathrm{z,int}(s) &= \delta_\mathrm{down,long}(s) \cdot \delta_\mathrm{down,lat}(s),  \\
	\label{eq:cz1}	\delta_\mathrm{down,long}(s) &= \mathcal{M_\mathrm{3}}(t_\mathrm{gap,rel}(s)), \\
	\label{eq:cz2}	\delta_\mathrm{down,lat}(s) &= \mathcal{M_\mathrm{4}}(y_\mathrm{gap,rel}(s)),
\end{alignat}
where \(\delta_\mathrm{down,long}\) and \(\delta_\mathrm{down,lat}\) are reduction factors fitted with \gls{nn} techniques from \cite{guerrero2020aerodynamic,dominy1990influence,ravelli2021aerodynamics}, alike the drag factors. The curves are shown in \Cref{fig:DownReduction}. Similarly to  \(\delta_\mathrm{drag,long}\),  \(\delta_\mathrm{down,long}\) reaches maximum reduction at \(t_\mathrm{gap,rel}(s)=\qty{0.2}{\second}\). \Cref{fig:DragReduction,fig:DownReduction} show the differences in lateral recovery between drag and downforce. It arises from component-specific aerodynamics. Downforce is dominated by underfloor, front wing, and rear wing midspan flows, which regain clear airflow with minimal lateral gap. Conversely, drag remains sensitive to wheel wake effects at larger offsets, as wheels disrupt airflow across a broader lateral range. Thus,  \(\delta_\mathrm{down,lat}\) drops rapidly with small offsets, while  \(\delta_\mathrm{drag,lat}\) requires larger offsets to diminish. This rapid downforce recovery at small lateral offsets enables competitive cornering maneuvers, where drivers minimize lateral displacement to retain grip, while exploiting drag reduction on straights.

Reduced downforce decreases vertical load,  which scales the maximum friction force via the performance envelope. The elliptical constraints’ semi-axes \(F_\mathrm{lat,max}\) and \(F_\mathrm{long,max}\) are thus scaled by \(1	- C_\mathrm{z,int} \), resulting in the following equations:  
\begin{alignat}{2}
	\left(\frac{F_\mathrm{lat}(s)}{F_\mathrm{lat,max}}\right)^2 + \left(\frac{F_\mathrm{long,acc}(s)}{F_\mathrm{long,max,acc}}\right)^2 &\leq (1 - C_{z,\mathrm{int}}(s))^2, \label{eq:longacc}\\
	\left(\frac{F_\mathrm{lat}(s)}{F_\mathrm{lat,max}}\right)^2 + \left(\frac{F_\mathrm{long,dec}(s)}{F_\mathrm{long,max,dec}}\right)^2 &\leq (1 - C_{z,\mathrm{int}}(s))^2. \label{eq:longdec}
\end{alignat}

\subsection{Optimal control problem formulation}
We now formulate the optimal control problem for each agent. After presenting the continuous formulation, we derive the discretized form.  
\begin{problem} \label{ocpsa}
	The \gls{ocp} for agent $i$ is to
	\begin{equation}
		\mathrm{minimize}_{P_{\mathrm{f},i}, P_{\mathrm{k},i}, P_{\mathrm{brk},i}, P_{\mathrm{K2ES},i},  a_{\mathrm{lat},i}} \quad J_{i}(s)
	\end{equation}
	subject to the following constraints: 
	\begin{alignat}{2}
		\mathrm{States:} &\quad \eqref{eq:spaceconv}, \nonumber \\
		\mathrm{Power \ unit:}  &\quad \eqref{eq:pp},\eqref{eq:pg},\eqref{eq:pu},\eqref{eq:pe},\eqref{eq:pkdc},\eqref{eq:constr}, \nonumber \\
		\quad &\quad \eqref{eq:eb},\eqref{eq:pb},\eqref{eq:pi},\nonumber\\
		\mathrm{External \ powers:} &\quad \eqref{eq:pext},\eqref{eq:paero},\eqref{eq:proll},\eqref{eq:pslope},\eqref{eq:paeroint}, \nonumber \\
		\mathrm{Performance \ envelope:} &\quad \eqref{eq:flongdef},\eqref{eq:flatdef},\eqref{eq:acclat},\eqref{eq:perfenv},\eqref{eq:flong}, \eqref{eq:flongacc}, \nonumber \\ 
		\quad &\quad \eqref{eq:flongdec}, \eqref{eq:longacc},\eqref{eq:longdec}, \nonumber\\
		\mathrm{Interaction \ constraints:} &\quad \eqref{eq:aeroint},\eqref{eq:cx}, \eqref{eq:cx1}, \eqref{eq:cx2}, \eqref{eq:cz},\eqref{eq:cz1}, \eqref{eq:cz2},  \nonumber \\
		\mathrm{Collision \ avoidance:} &\quad \eqref{eq:collision },\nonumber \\
		\mathrm{Trajectory:} &\quad  \eqref{eq:tracklim}.\nonumber 
	\end{alignat} 
\end{problem}	
The \gls{ocp} is now converted into a \gls{nlp} through multiple shooting method and Euler forward integration scheme. In particular, the track is discretized in $N$ steps denoted by $k$: 
\begin{equation}
	s^{k}\in[0,S], \quad k\in\{1,\dots,N\},
\end{equation}
where the number of steps $N$ ranges between 430 and 600, depending on the length of the track. The step size
\begin{equation}
\Delta s^{k} = s^{k+1} - s^{k}, \quad k\in\{1,\dots,N-1\},
\end{equation}
is not equally spaced: The mesh is finer in the corners and coarser in the straights. This allows to capture the steep state gradients of the corners while remaining computationally efficient.
The input and state vectors for step $k$ are: 
\begin{alignat}{2}
	\mathbf{u}_{i}^{k} &= \begin{bmatrix}
		P_{\mathrm{f},i}^{k} & P_{\mathrm{k},i}^{k} & P_{\mathrm{brk},i}^{k} & P_{\mathrm{K2ES},i}^{k} & a_{\mathrm{lat},i}^{k}  
	\end{bmatrix}, \\ 
	&\hspace{5cm} k \in \{1, \dots, N-1\}, \nonumber \\
	\mathbf{x}_{i}^{k} &= \begin{bmatrix}
		v_{i}^{k} & E_{\mathrm{f},i}^{k} & E_{\mathrm{b},i}^{k} & t_{i}^{k} & E_{\mathrm{K2ES},i}^{k} & y_{i}^\mathrm{k} & \varphi_{i}^{k}  
	\end{bmatrix}, \\ 
	&\hspace{5cm} k \in \{1, \dots, N\}. \nonumber
\end{alignat}
while the vectors for the entire lap are
\begin{alignat}{2}
	\mathbf{u} &= \begin{bmatrix}
	\mathbf{u}_{i}^\mathrm{1} & \dots & \mathbf{u}_{i}^{N-1} 
	\end{bmatrix}^\intercal, \\
	\mathbf{x} &= \begin{bmatrix}
		\mathbf{x}_{i}^\mathrm{1} & \dots &\mathbf{x}_{i}^{N} 
	\end{bmatrix}^\intercal.
\end{alignat}

\begin{problem}
	The \gls{nlp} resulting from the transcription of \cref{ocpsa} for the single agent case is  
	\begin{alignat*}{2}
		& \min_{\mathbf{x}_{i}, \mathbf{u}_{i}} J_{i}(\mathbf{x}_{i},\mathbf{u}_{i}, \mathbf{x}_{- i})\\
		& \text{subject to:} \\
		& \quad \quad \mathbf{g}_{i}(\mathbf{x}_{i},\mathbf{u}_{i}, \mathbf{x}_{- i}) \le 0, \\
		& \quad \quad \mathbf{h}_{i}(\mathbf{x}_{i},\mathbf{u}_{i}, \mathbf{x}_{- i}) = 0,\\
	\end{alignat*}  
	\(\mathbf{g}_{i}\) and \(\mathbf{h}_{i}\) are a vectorized collection of all the inequality and equality constraints of \cref{ocpsa} respectively.  \(\mathbf{h}_{i}\) includes the continuity constraints resulting from multiple shooting formulation.
\end{problem}

So far, we have been expressing the cost function with a general notation $J_{i}$. In \cref{sec:method}, the same convention will persist. 
In single-agent formulations, the objective is to minimize the agent's own lap time, independently of other agents, reading
\begin{equation}
	J_{i}(s) = t_{i}(S) = \int_{0}^{S}\frac{1}{v_{\mathrm{c},i}(s)}\mathrm{d}s. 
\end{equation}
According to the change of variables of \cref{eq:dds}, the integrand involves $v_{\mathrm{c},i}$. The physical interpretation is that the progression on the circuit is given by the velocity along the centerline, and not by the absolute velocity resulting from lateral deviations. The discretized formulation is 
\begin{equation}
	J_{i}(\mathbf{x}_{i}) = t_{i}^{N} = \sum_{k=1}^{N-1}\frac{\Delta s^{k}}{v_{c,i}^k}.
\end{equation} However, in multi-agent scenarios, the objective function might incorporate the lap time of the competing agents, depending on the game formulation. 

We acknowledge that other cost functions might be implemented, such as those focusing on battery energy consumption or fuel efficiency. These functions would be highly relevant in other contexts, such as series cars, where the trade-off between efficiency and performance is of primary importance. Nevertheless, due to the nature of motorsport, where strategic interactions arise in competitive scenarios, the objective remains to drive as fast as possible, minimizing lap time.

%% file: figures/TrajectoryFrame/TrajectoryFrame.tex
	
	\begin{tikzpicture}
		\tikzset{>=stealth}

		\definecolor{lightblue}{rgb}{0.4,0.8,1}
		\definecolor{lightgray}{gray}{0.8}

		\draw[black, thick, ->] plot [smooth, tension=1] coordinates {(0,0) (1.5,0.2) (3,-0.2) (4.5,0.5) (6,0)} node[anchor=south, yshift=0.1cm] {$s$};

		\coordinate (lastpoint) at (6,0);
		\coordinate (firstpoint) at (0,0);

		\draw[dashed] (firstpoint) -- ++(-0.75,-0.25);

		\draw[->, black] (lastpoint) -- ++(-0.67,-0.9) node[midway, anchor=east] {$y$};
		
		\filldraw[black] (5.25,-1) circle (3.5pt) node[anchor=east, xshift=-0.2cm] {Vehicle position};

		\draw[dashed] (lastpoint) -- ++(1,-0.75);

		\draw[->, black] (lastpoint) ++(-0.75,-1) -- ++(1,-1.4) node[anchor=west] {$v$};

		\draw[->, black] (lastpoint) ++(-0.75,-1) -- ++(-1.4,-1) node[anchor=west, , xshift=0.2cm] {$a_\mathrm{lat}$};

		\draw[dashed] (lastpoint) ++(-0.75,-1) -- ++(1,-0.75);

		\draw[->, black] (lastpoint) ++(0.3,-1.75) arc[start angle=-40, end angle=-60, radius=1.1] node[ anchor=west, xshift=0.05cm,yshift=-0.05cm] {$\varphi$};

		\draw[->] (1.6,-1.6) -- ++(0.2,1.75) node[midway, anchor=east] {$r_\mathrm{c}$};
		
		\filldraw[fill=white, draw=black, thick] (1.6,-1.6) circle (1pt);

		\node at (6, 0) [above, yshift=0.6cm, xshift=-1.5cm] {Track centerline};
		
	\end{tikzpicture}

%% file: figures/RelGap/RelGap.tex
	\centering
	\begin{tikzpicture}
		\tikzset{>=stealth}
		\draw[thick, ->,blue,dashed] (-8,-1.3) to[curve through={ (-5.95,-0.4) (-2,-0.5) }] (-1.55,-0.2);
		\draw[thick, ->,red,dashed] (-8, 0) to[curve through={  (-5.5,0.5) (-2,1)}] (-1.55,1.5);
		
		\draw[black, dashed, thick] (-7.85,-0.82) -- (-7.5,-0.3);
		\draw[black, dashed, thick] (-1.9,0.5355) -- (-1.1,0.51);
	
		\draw[thick, ->] (-7.5,-0.3) to[curve through={ (-5.95,0.12) (-5.75,0) (-5.6,-0.08) (-4,-0.6) (-2.1,0.5355) (-2,0.5368) }] (-1.55,0.52);

		\filldraw[red,draw=black] (-5.5,0.5) circle (3pt) node[anchor=south, shift={(0,0.1)}] { \parbox{2cm}{ $ t_\mathrm{A}(s_\mathrm{1})=\qty{7}{\second}$ \\ \centering $A$ }}; 
		\draw[<-,red] (-5.55,0.4) --  (-5.75,0) node[anchor=south, shift={(0.8,-0.2)}] {$y_\mathrm{A}(s_\mathrm{1})$};

		\filldraw[blue,draw=black] (-5.95,-0.4) circle (3pt) node[anchor=north, shift={(0,-0.2)}] {\parbox{2cm}{\centering $B$ \\  $ t_\mathrm{B}(s_\mathrm{1})=\qty{8}{\second}$}}; 
		\draw[->,blue] (-5.75,0) -- (-5.9,-0.3) node[anchor=south, shift={(-0.6,-0.1)}] {$y_\mathrm{B}(s_\mathrm{1})$};

    	\filldraw[red,draw=black] (-2,1) circle (3pt) node[anchor=north, shift={(0,1)}] {\parbox{2cm}{\centering $ t_\mathrm{A}(s_\mathrm{2})=\qty{8}{\second}$ \\ $A$}}; 
    	\draw[<-,red] (-2,0.89) -- (-2,0.5368) node[anchor=east, shift={(-0.3,0.45)}] {$y_\mathrm{A}(s_\mathrm{2})$};

		\filldraw[blue,draw=black] (-2,-0.5) circle (3pt) node[anchor=south, shift={(0,-1.1)}]   {\parbox{2cm}{\centering $B$ \\  \centering $t_\mathrm{B}(s_\mathrm{2})=\qty{9}{\second}$}};  
		\draw[->,blue] (-2,0.5368) -- (-2,-0.4) node[anchor=west, shift={(0,0.5)}] {$y_\mathrm{B}(s_\mathrm{2})$};		

		\draw[->,black] (-4.23,-0.63) -- (-4.23,-1.1) node[anchor=north, shift={(0,0.05)}] {$y$};
	         \draw[] (-1.55,0.52) node[anchor=south, shift={(0.05,0)}] {$s$};

  \end{tikzpicture}

%% file: figures/DragReduction/DragReduction.tex
		\begin{tikzpicture}[trim axis right] 
			\ifdefined\isMainDocument
			\def\datapath{figures/DragReduction/}
			\def\plotwidth{\columnwidth}%
			\else
			\def\datapath{}
			\def\plotwidth{7cm}
			\fi
			
			\def\yshift{-0.18cm}%
			\def\plotheight{5cm}%
			\def\verticalOffset{0cm}		
			\def\xmaxLong{2}	
			\def\ymaxLong{0.45}
			\def\xmaxLat{2.5}	
			\def\ymaxLat{1.1}
					
			\begin{axis}[			
				width=\plotwidth,
				height=\plotheight,
				at={(0,0)},
				xmin=-0.2,
				xmax=\xmaxLong,
				xlabel style={font=\color{white!15!black}},
				xlabel={$t_{\mathrm{gap,rel}}$},
				x unit =\unit{\second},
				ymin=-0.01,
				ymax=\ymaxLong,
				ylabel style={font=\color{white!15!black}},
				ylabel={$\delta_{\mathrm{drag,long}}$},
				ylabel style={font=\color{white!15!black}},
				y unit =\unit{\relax-},			
				axis background/.style={fill=white},
				xmajorgrids,
				ymajorgrids,
				legend style={legend cell align=left, align=left, draw=white!15!black, at={(axis cs:\xmaxLong,\ymaxLong)}, anchor=north east}
				]
				\addplot [color=black, line width=1pt]
				table[]{\datapath DragLongitudinalplots-3.tsv};
				\addlegendentry{Model}
				
				\addplot[only marks, line width=1pt, mark=o, mark options={}, mark size=3pt, draw=gray, fill=gray] table[]{\datapath DragLongitudinalplots-1.tsv};
				\addlegendentry{\cite{guerrero2020aerodynamic}}
				
				\addplot[only marks, line width=1pt, mark=x, mark options={}, mark size=3pt, draw=gray, fill=gray] table[]{\datapath DragLongitudinalplots-2.tsv};
				\addlegendentry{\cite{ravelli2021aerodynamics}}
				
			\end{axis}
			
			\begin{axis}[
				width=\plotwidth,
				height=\plotheight,
				at={(0, -\plotheight - \verticalOffset)},
				xmin=-2.5,
				xmax=\xmaxLat,
				xlabel style={font=\color{white!15!black}},
				xlabel={$\frac{y_{\mathrm{gap,rel}}}{\mathrm{CarWidth}}$},
				x unit =\unit{\relax-},
				ymin=-0.01,
				ymax=\ymaxLat,
				ylabel style={font=\color{white!15!black}},
				ylabel={$\delta_{\mathrm{drag},\mathrm{lat}}$},
				ylabel style={font=\color{white!15!black}},
				y unit =\unit{\relax-},
				axis background/.style={fill=white},
				xmajorgrids,
				ymajorgrids,
				legend style={legend cell align=left, align=left, draw=white!15!black, at={(axis cs:\xmaxLat,\ymaxLat)}, anchor=north east}
				]
							
				\addplot [color=black, line width=1pt]
				table[]{\datapath DragLateralplots-2.tsv};
				\addlegendentry{Model}
				
				\addplot[only marks, line width=1pt, mark=diamond, mark options={}, mark size=3pt, draw=gray, fill=gray] table[]{\datapath DragLateralplots-1.tsv};
				\addlegendentry{\cite{dominy1990influence}}
				
			\end{axis}

		\end{tikzpicture}

%% file: figures/DownforceReduction/DownforceReduction.tex
	
	\begin{tikzpicture}[trim axis right] 
		\ifdefined\isMainDocument
		\def\datapath{figures/DownforceReduction/}
		\def\plotwidth{\columnwidth}%
		\else
		\def\datapath{}
		\def\plotwidth{7cm}
		\fi
		
		\def\yshift{-0.18cm}%
		\def\plotheight{5cm}%
		\def\verticalOffset{0cm}		
		\def\xmaxLong{2}	
		\def\ymaxLong{0.65}
		\def\xmaxLat{2.5}	
		\def\ymaxLat{1.1}
							
		\begin{axis}[			
			width=\plotwidth,
			height=\plotheight,
			at={(0,0)},
			xmin=-0.2,
			xmax=\xmaxLong,
			xlabel style={font=\color{white!15!black}},
			xlabel={$t_{\mathrm{gap,rel}}$},
			x unit=\unit{\second},
			ymin=-0.01,
			ymax=\ymaxLong,
			ylabel style={font=\color{white!15!black}},
			ylabel={$\delta_{\mathrm{down,long}}$},
			ylabel style={font=\color{white!15!black}},
			y unit =\unit{\relax-},
			axis background/.style={fill=white},
			xmajorgrids,
			ymajorgrids,
			legend style={legend cell align=left, align=left, draw=white!15!black, at={(axis cs:\xmaxLong,\ymaxLong)}, anchor=north east}
			]
			\addplot [color=black, line width=1pt]
				table[]{\datapath DownLongitudinalplots-3.tsv};
				\addlegendentry{Model}
				
				\addplot[only marks, line width=1pt, mark=o, mark options={}, mark size=3pt, draw=gray, fill=gray] table[]{\datapath DownLongitudinalplots-1.tsv};
				\addlegendentry{\cite{guerrero2020aerodynamic}}
				
				\addplot[only marks, line width=1pt, mark=x, mark options={}, mark size=3pt, draw=gray, fill=gray] table[]{\datapath DownLongitudinalplots-2.tsv};
				\addlegendentry{\cite{ravelli2021aerodynamics}}
			
		\end{axis}
		
		\begin{axis}[
			width=\plotwidth,
			height=\plotheight,
			at={(0, -\plotheight - \verticalOffset)},
			xmin=-2.5,
			xmax=\xmaxLat,
			xlabel style={font=\color{white!15!black}},
			xlabel={$\frac{y_{\mathrm{gap,rel}}}{\mathrm{CarWidth}}$},
			x unit =\unit{\relax-},
			ymin=-0.01,
			ymax=\ymaxLat,
			ylabel style={font=\color{white!15!black}},
			ylabel={$\delta_{\mathrm{down},\mathrm{lat}}$},
			ylabel style={font=\color{white!15!black}},
			y unit =\unit{\relax-},
			axis background/.style={fill=white},
			xmajorgrids,
			ymajorgrids,
			legend style={legend cell align=left, align=left, draw=white!15!black, at={(axis cs:\xmaxLat,\ymaxLat)}, anchor=north east}
			]
					
			\addplot [color=black, line width=1pt]
			table[]{\datapath DownLateralplots-2.tsv};
			\addlegendentry{Model}
			
			\addplot[only marks, line width=1pt, mark=diamond, mark options={}, mark size=3pt, draw=gray, fill=gray] table[]{\datapath DownLateralplots-1.tsv};
			\addlegendentry{\cite{dominy1990influence}}
			
		\end{axis}
	\end{tikzpicture}

%% file: chapters/S3_Methodology.tex
\section{Methodology}\label{sec:method}
In this section, the considered game-theoretic formulations are explained in conjunction with their mathematical properties. Afterwards, we present the algorithm to find a better local Stackelberg solution.

\subsection{Dynamic games}
Dynamic games are a promising approach to explore multi-agent interactions within optimization frameworks. A distinctive feature of games is the involvement of multiple players. Depending on their objectives, the degree of conflict between them, and the game under consideration, their behavior can substantially change. In contrast, standard optimization problems only allow one or multiple players to act in full cooperation to minimize a common objective. In this study, we focus on the interaction between two cars in a \gls{f1} race, whose goals are never completely cooperative. 

Game theory includes several types of games, each characterized by unique properties and equilibrium points. Among them, Stackelberg and Nash games are particularly suited to capture the interplays in motorsport. Stackelberg games exhibit a leader-follower structure, where one agent is subject to the decisions of the other. For instance, this can happen when an experienced pilot is defending or attacking a position against a rookie. By overtaking or blocking maneuvers, the trajectory of the other vehicle can be forced to deviate from the desired one. On the other hand, in Nash games, the agents have equal decision power and are thus more balanced. Using the same example, two equally experienced pilots are fighting for a position.

Dynamic games are in close relation to optimal control theory \cite{geering2007optimal, ungureanu2018pareto}, the tools of which can be leveraged to find numerical solutions. For the games considered, we can find formulations in the literature that can be reduced to \glspl{nlp} \cite{pilecka2012combined,bacsar1998dynamic,dempe2020bilevel}. In this perspective, Stackelberg and Nash games share the same problem setup, and they only differ in the formulation of the \gls{nlp}. We will explore how the different game setups affect the resulting strategies.

Our focus is to study the physical interaction on a single lap. To this end, we consider two identical agents. We will address them as $A$ and $B$, and they are interchangeable in the various formulations. Additionally, as previously stated, we use the index $i$ to indicate ``agent $i$'' and $- i$ to indicate ``not agent $i$''.

\subsubsection{Stackelberg game}
Stackelberg games have a hierarchy in the form of leader-follower relationship. The leader takes an action, with the awareness that the follower will optimally respond to it. Usually, this is captured by a sequential game, meaning that the leader publishes its decision, and only afterwards, the follower makes its move. In our case, since we are dealing with a dynamic game, we lose the sequentiality of the actions, resulting in a simultaneous game.

Mathematically, the decision-making process of a Stackelberg game can be seen as a two-level optimization. The leader optimization problem is constrained by the follower optimal response as shown in \Cref{prob:bilevel}. 

\begin{problem}\label{prob:bilevel}
The bilevel program capturing the dynamic Stackelberg game is
\begin{alignat*}{2}
 & \min_{\mathbf{x}_{\mathrm{L}}, \mathbf{u}_{\mathrm{L}}} J_{\mathrm{L}}(\mathbf{x}_{\mathrm{L}},\mathbf{u}_{\mathrm{L}}, \mathbf{x}_{\mathrm{F}})\\
 & \text{subject to:} \\
		& \quad \quad \mathbf{g}_{\mathrm{L}}(\mathbf{x}_{\mathrm{L}},\mathbf{u}_{\mathrm{L}}, \mathbf{x}_{\mathrm{F}}) \le 0, \\
		& \quad \quad \mathbf{h}_{\mathrm{L}}(\mathbf{x}_{\mathrm{L}},\mathbf{u}_{\mathrm{L}}, \mathbf{x}_{\mathrm{F}}) = 0,\\
		& \quad \quad \{\mathbf{x}_{\mathrm{F}},\mathbf{u}_{\mathrm{F}}\} = \arg\min_{\mathbf{x}_{\mathrm{F}}, \mathbf{u}_{\mathrm{F}}}  J_{\mathrm{F}}(\mathbf{x}_{\mathrm{F}},\mathbf{u}_{\mathrm{F}}, \mathbf{x}_{\mathrm{L}})\\
		& \quad \quad \text{subject to:} \\
		& \quad \quad \quad \quad \mathbf{g}_{\mathrm{F}}(\mathbf{x}_{\mathrm{F}},\mathbf{u}_{\mathrm{F}}, \mathbf{x}_{\mathrm{L}}) \le 0, \\
		& \quad \quad \quad \quad \mathbf{h}_{\mathrm{F}}(\mathbf{x}_{\mathrm{F}},\mathbf{u}_{\mathrm{F}}, \mathbf{x}_{\mathrm{L}}) = 0,
\end{alignat*}
where $\mathrm{L}$ stands for the leader and $\mathrm{F}$ for the follower.
\end{problem}

However, solving this kind of problem is generally challenging. One possibility is to replace the low-level problem with its closed-form solution (if it exists). Another common approach is to reformulate the low-level program with the \gls{kkt} conditions as in \cite{schwarting2019social,burger2022interaction}. In both cases, we obtain a single-level \gls{nlp}, which can be tackled by off-the-shelf solvers. Given the complexity of our system and its nonlinearities, we discard the first option. We opt for the \gls{kkt}-based numerical scheme, based on the results of \cite{fieni2024game}, showing its efficiency and reliability.

\Cref{def:kkt} summarizes the \gls{kkt} conditions, where we also introduce a short-hand notation. 

\begin{Def}\label{def:kkt}
For an optimization problem of the form 
\begin{alignat*}{2}
 & \min_{\mathbf{x}_{i}, \mathbf{u}_{i}} J_{i}(\mathbf{x}_{i},\mathbf{u}_{i}, \mathbf{x}_{- i})\\
 & \text{subject to:} \\
		& \quad \quad \mathbf{g}_{i}(\mathbf{x}_{i},\mathbf{u}_{i}, \mathbf{x}_{- i}) \le 0, \\
		& \quad \quad \mathbf{h}_{i}(\mathbf{x}_{i},\mathbf{u}_{i}, \mathbf{x}_{- i}) = 0,\\
\end{alignat*}
with the Lagrangian 
\begin{align}\label{lagrangian}
L_{i}(\mathbf{x}_{i},\mathbf{u}_{i}, \mathbf{x}_{- i}, \boldsymbol{\lambda}_{i}, \boldsymbol{\mu}_{i}) = & J_{i}(\mathbf{x}_{i},\mathbf{u}_{i}, \mathbf{x}_{- i}) \nonumber\\
& +\boldsymbol{\lambda}^{\intercal}_{i}\cdot \mathbf{h}_{i}(\mathbf{x}_{i},\mathbf{u}_{i}, \mathbf{x}_{- i})  \nonumber\\
& +\boldsymbol{\mu}^{\intercal}_{i}\cdot \mathbf{g}_{i}(\mathbf{x}_{i},\mathbf{u}_{i}, \mathbf{x}_{- i}),
\end{align}
the \gls{kkt} conditions read
\begin{alignat}{2}
		 &\quad \quad \nabla_{\mathbf{x}_{i},\mathbf{u}_{i}}L_{i}(\mathbf{x}_{i},\mathbf{u}_{i}, \mathbf{x}_{- i}, \boldsymbol{\lambda}_{i}, \boldsymbol{\mu}_{i})  = 0, \label{eq:stationarity}\\
		 &\quad \quad \mathbf{g}_{i}(\mathbf{x}_{i},\mathbf{u}_{i}, \mathbf{x}_{- i})  \le 0, \label{eq:primal1}\\
		 &\quad \quad \mathbf{h}_{i}(\mathbf{x}_{i},\mathbf{u}_{i}, \mathbf{x}_{- i})  = 0, \label{eq:primal2}\\
		 &\quad \quad \boldsymbol{\mu}_{i} \ge 0, \label{eq:dual}\\
		 &\quad \quad \mu_{i,j}\cdot {g}_{i,j}(\mathbf{x}_{i},\mathbf{u}_{i}, \mathbf{x}_{- i})   = 0, \quad j\in\{1,\dots,m\}, \label{eq:slackness}
\end{alignat}
where \cref{eq:stationarity} is the stationarity condition, \cref{eq:primal1,eq:primal2} are the primal feasibility, \cref{eq:dual} is the dual feasibility, \cref{eq:slackness} is the complementary slackness and $m$ is the number of inequality constraints.\\
We define the compact form of the \gls{kkt} conditions as 
\begin{equation}
\mathcal{KKT}_{i}(\mathbf{x}_{i},\mathbf{u}_{i}, \mathbf{x}_{- i},\boldsymbol{\lambda}_{i},\boldsymbol{\mu}_{i}).
\end{equation}
\end{Def}

To avoid the complications of a \gls{mpcc}, we relax the constraints of \cref{eq:slackness} with the Scholtes' relaxation scheme \cite{scholtes2001convergence} 
\begin{equation}
\mu_{i,j}\cdot {g}_{i,j}(\mathbf{x}_{i},\mathbf{u}_{i}, \mathbf{x}_{-i})  \ge -\varepsilon, \quad j\in\{1,\dots,m\},
\end{equation}
where $\varepsilon\ge 0$, recovering the \gls{mfcq} \cite{hoheisel2013theoretical}. Further details can be found in \cite{fieni2024game}.

Note that the variable $\mathbf{x}_{- i}$ in \cref{def:kkt} is not an optimization variable of the problem. For this reason, it is treated as a constant during the computation of the Lagrangian's gradient. However, we keep it as a placeholder, because in the subsequent Stackelberg game, it is indeed an optimization variable of the whole problem. The Stackelberg game reformulation is presented in \Cref{prob:stackelbergRef}. 

\begin{problem}\label{prob:stackelbergRef}
The \gls{kkt}-based reformulation of the Stackelberg game reads
\begin{alignat*}{2}
 & \min_{\mathbf{x}_{\mathrm{L}}, \mathbf{u}_{\mathrm{L}}, \mathbf{x}_{\mathrm{F}}, \mathbf{u}_{\mathrm{F}}, \boldsymbol{\lambda}_{\mathrm{F}}, \boldsymbol{\mu}_{\mathrm{F}}} J_{\mathrm{L}}(\mathbf{x}_{\mathrm{L}},\mathbf{u}_{\mathrm{L}}, \mathbf{x}_{\mathrm{F}}) + J_{\mathrm{F}}(\mathbf{x}_{\mathrm{F}},\mathbf{u}_{\mathrm{F}}, \mathbf{x}_{\mathrm{L}})\\
 & \text{subject to:} \\
		& \quad \quad \mathbf{g}_{\mathrm{L}}(\mathbf{x}_{\mathrm{L}},\mathbf{u}_{\mathrm{L}}, \mathbf{x}_{\mathrm{F}}) \le 0, \\
		& \quad \quad \mathbf{h}_{\mathrm{L}}(\mathbf{x}_{\mathrm{L}},\mathbf{u}_{\mathrm{L}}, \mathbf{x}_{\mathrm{F}}) = 0,\\
		& \quad \quad \mathcal{KKT}_{\mathrm{F}}(\mathbf{x}_{\mathrm{F}},\mathbf{u}_{\mathrm{F}}, \mathbf{x}_{\mathrm{L}},\boldsymbol{\lambda}_{\mathrm{F}},\boldsymbol{\mu}_{\mathrm{F}}),
\end{alignat*}
which is a single-level nonlinear program.
\end{problem}
Note that in this step, we added the follower's cost to the objective. The \gls{kkt} conditions enforce to solve for stationary points of the low-level program. However, since they are first-order conditions, further analysis would be necessary to distinguish between local maxima, minima, or saddle points. This requires additional second-order constraints or an analysis in post-processing. As suggested in \cite{schwarting2019social}, the cost of the low-level program can be added in the final reformulation. This ensures that the solver will search for a local minimum.

In \cref{def:stackelberg}, we introduce the short-hand notation for the (locally) optimal solution of \Cref{prob:stackelbergRef}. 
\begin{Def}\label{def:stackelberg}
The optimal solution of \Cref{prob:stackelbergRef} in compact form is defined as 
\begin{equation}
  \mathcal{SG}(i,-i) := \{J_{i,\mathrm{L}}^\star, \mathbf{x}_{i,\mathrm{L}}^\star,\mathbf{u}_{i,\mathrm{L}}^\star, J_{-i,\mathrm{F}}^\star, \mathbf{x}_{-i,\mathrm{F}}^\star,\mathbf{u}_{-i,\mathrm{F}}^\star\}.
\end{equation}
For instance, given two agents $A$ and $B$, $\mathcal{SG}(A,B)$ represents the solution of the Stackelberg game with $A$ as leader and $B$ as follower.
\end{Def}

\subsubsection{Nash game}
Unlike the previous case, Nash games do not exhibit a hierarchical structure. Thus, the logic in the decision-making process is different to that in the Stackelberg game. The interactions are symmetric, since there are no defined roles such as leader and follower. This means that there are no roles distinguishing between the two agents.

\Cref{prob:nash} depicts the structure of the related mathematical optimization. Searching for a Nash equilibrium is equivalent to solve simultaneously $n$ interdependent optimization problems, one for each agent. 
\begin{problem}\label{prob:nash}
The optimization problems describing the Nash game are
\begin{alignat*}{2}
 & \min_{\mathbf{x}_{i}, \mathbf{u}_{i}} J_{i}(\mathbf{x}_{i},\mathbf{u}_{i}, \mathbf{x}_{- i})\\
 & \text{subject to:} \\
		& \quad \quad \mathbf{g}_{i}(\mathbf{x}_{i},\mathbf{u}_{i}, \mathbf{x}_{- i}) \le 0, \\
		& \quad \quad \mathbf{h}_{i}(\mathbf{x}_{i},\mathbf{u}_{i}, \mathbf{x}_{- i}) = 0,\\
		& & \quad \forall i\in\{1,\dots,n\},
\end{alignat*}
where $n$ is the number of agents.
\end{problem}

The typical solution approach is the \gls{ibr}. In this approach, one problem at a time is solved, keeping the variables of the other agents $-i$ constant. After updating the new variables of agent $i$, the same optimization is carried out for the next agent: This procedure is iteratively repeated until convergence. If the algorithm converges, then a Nash equilibrium is found. However, drawbacks of iterative schemes, such as oscillations, limit its use in practice, particularly for medium- and large-scale nonlinear programs.

As an alternative, in \cite{schwarting2019social}, a \gls{kkt}-based solution approach is employed to solve the $n$ dependent optimization problems. As done for the reformulation of the low-level program in the Stackelberg game, we replace each optimization problem with its \gls{kkt} conditions. The objective is then added to the total cost, and the final formulation is presented in \Cref{prob:nashRef}.
\begin{problem}\label{prob:nashRef}
The \gls{kkt}-based reformulation of the Nash game reads
\begin{alignat*}{2}
 & \min_{\mathbf{x}, \mathbf{u}, \boldsymbol{\lambda}, \boldsymbol{\mu}} \sum_{i=1}^{n} J_{i}(\mathbf{x}_{i},\mathbf{u}_{i}, \mathbf{x}_{-i})\\
 & \text{subject to:} \\
		& \quad \quad \mathcal{KKT}_{1}(\mathbf{x}_{1},\mathbf{u}_{1}, \mathbf{x}_{-1},\boldsymbol{\lambda}_{1},\boldsymbol{\mu}_{1}),\\
		& \quad \quad \vdots \\
		& \quad \quad \mathcal{KKT}_{n}(\mathbf{x}_{n},\mathbf{u}_{n}, \mathbf{x}_{-n},\boldsymbol{\lambda}_{n},\boldsymbol{\mu}_{n}),
\end{alignat*}
which is a single nonlinear program, with $n$ being the number of agents. The vectors $\mathbf{x}$, $\mathbf{u}$, $\boldsymbol{\lambda}$, and $\boldsymbol{\mu}$, respectively, summarize the states, input and costate vectors of every agent.
\end{problem}
This formulation allows to solve all \glspl{nlp} in one single optimization problem, without employing iterative schemes.

Comparing the resulting games, we observe that they are closely related, making it easy to switch between them. It is easy to convert a Stackelberg game into a Nash game and vice versa, since only the KKT reformulations alter the decision-making logic of a particular game. \Cref{def:nash} introduces the short-hand notation for the solution of the Nash game.
\begin{Def}\label{def:nash}
The optimal solution of \Cref{prob:nashRef}  for two agents in compact form is defined as 
\begin{equation}
 \mathcal{NG}(i,-i) := \{J_{i,\mathrm{N}}^\star, \mathbf{x}_{i,\mathrm{N}}^\star,\mathbf{u}_{i,\mathrm{N}}^\star, J_{-i,\mathrm{N}}^\star, \mathbf{x}_{-i,\mathrm{N}}^\star,\mathbf{u}_{-i,\mathrm{N}}^\star\}.
\end{equation}
For instance, given two agents $A$ and $B$, $\mathcal{NG}(A,B)$ represents the solution of the Nash game.
\end{Def}

\ref{sec:appendix_1} shows how the Nash and Stackelberg formulations can be combined to describe a single-leader-multi-follower game.

\subsubsection{Finding a better Stackelberg solution}\label{sec:stackAlg}
Despite the conception of equilibrium, a Nash equilibrium (if it exists) might not be the best solution for one or all agents. Indeed, if one agent is the leader of a game, it can take decisions which are better for itself (and they could even be better for the follower). The prisoner's dilemma \cite{tucker1950two} provides an example. In the ``classical'' Nash setup, if both prisoners cooperate, they will both face a reduced sentence. However, if one testifies before the other, i.e., he/she takes a leading role and walks free, the other one faces the full sentence. This reasoning gives us an intuition that \textit{the leader can always perform at least as good as its Nash solution} \cite{bacsar1998dynamic,simaan1977equilibrium}. 

With this concept in mind, the Stackelberg and Nash solutions can be compared to each other. Let us consider 3 games with the resulting costs of the agents $A$ and $B$:
\begin{itemize}
\item Nash solution: $J_{A,\mathrm{N}}$, $J_{B,\mathrm{N}}$
\item Stackelberg solution with leader $A$: $J_{A,\mathrm{L}}$, $J_{B,\mathrm{F}}$
\item Stackelberg solution with leader $B$: $J_{B,\mathrm{L}}$, $J_{A,\mathrm{F}}$
\end{itemize}
As introduced earlier, the leader can always perform at least as good as its Nash solution, which translates to 
\begin{align}
J_{A,\mathrm{L}}&\le J_{A,\mathrm{N}},\label{EqLeader1}\\
J_{B,\mathrm{L}}&\le J_{B,\mathrm{N}}. \label{EqLeader2}
\end{align}
In \cite{simaan1973stackelberg}, the agents' costs in Nash and Stackelberg games are studied. A Stackelberg solution is defined as \textit{dominant} if both the leader and the follower achieve better outcomes compared to the Nash equilibrium. Conversely, if only the leader improves its cost, the solution is called \textit{non-dominant}. It is worth noting that in non-zero-sum games, a reduction in the leader cost is not necessarily accompanied by a corresponding increase in the follower cost.

However, a downside of \gls{nlp} solvers is that they do not guarantee finding the global optimum. Solving the reformulated Stackelberg game may lead to a higher leader cost compared to the Nash solution. To address this, we leverage the aforementioned mathematical property to obtain a better Stackelberg solution. Given a Nash solution, we can always find a policy by solving the corresponding Stackelberg game that either improves or maintains the leader's cost. In practice, the Nash game's resulting cost serves as an upper bound for the leader's objective. \Cref{alg:computeSol} implements the discussed method. Although we cannot conclude anything about global optimality, we can ensure a certain degree of comparability with the Nash game. From \Cref{alg:computeSol}, we get a set of solutions $\mathcal{T}$ composed by 3 comparable games: The Nash game and two Stackelberg games with $A$ as leader with $B$ as leader each, with the guarantee of an equal or better cost for the leader. 
\begin{algorithm}
\caption{Refined Stackelberg solutions}\label{alg:computeSol}
\begin{algorithmic}
\onehalfspacing
\State \textbf{Input:} Boundary conditions equal for all problems.
\Procedure{Compute set of solutions}{}
\State Find $\mathcal{NG}(A,B)$ by solving \cref{prob:nashRef}
\State $J_{A,\mathrm{N}}, J_{B,\mathrm{N}} \gets J_{A,\mathrm{N}}^{\star}, J_{B,\mathrm{N}}^{\star}$
\For{$i \in \{A,B\}$}
\State 1) Set leader's objective upper bound: 
\State \indent $J_{i,\mathrm{L}}(\mathbf{x}_{i,\mathrm{L}},\mathbf{u}_{i,\mathrm{L}}, \mathbf{x}_{i,\mathrm{F}}) \le J_{i,\mathrm{N}}$
\State 2) Add this bound to the set of inequality constraints \indent \indent of \cref{prob:stackelbergRef}
\State 3) Find $\mathcal{SG}(i,-i)$ of modified \cref{prob:stackelbergRef}
\EndFor 
\EndProcedure
\State \textbf{Result:} Solution set $\mathcal{T} = \{\mathcal{NG}(A,B), \mathcal{SG}(A,B), \mathcal{SG}(B,A)\}$
\end{algorithmic}
\end{algorithm}

This approach comes with the following advantages: The solution is guaranteed to be feasible, since we exploit a mathematical property of games. Additionally, the improvement is achieved without recursion or iterative processes, avoiding convergence and infeasibility issues. Since we are also interested in the Nash equilibrium, solving beforehand the Nash game to get the objectives' upper bounds is already part of the solution process.

The idea can be further extended to find dominant solutions, i.e., Stackelberg games where both leader and follower achieve a better cost than the Nash solution:
\begin{align}
J_{i,\mathrm{L}}&\le J_{i,\mathrm{N}},\\
J_{-i,\mathrm{F}}&\le J_{-i,\mathrm{N}}.
\end{align}
However, there is no guarantee that such solutions exist. Additionally, the same Stackelberg game can be solved with $i$ as leader and $-i$ as follower and vice versa. It becomes then important to distinguish between concurrent, nonconcurrent or stalemate solution \cite{basar1973relative,simaan1977equilibrium}, which are discussed in~\ref{app:equilibria}. For this characterization, an iterative process might be needed. However, this is not the focus of this paper.

\subsection{Computational details}
\Cref{prob:nashRef} corresponds to the largest \gls{nlp} formulation considered in this work. Typical problem sizes feature approximately 70000 optimization variables and 257000 constraints. The problems are parsed using CasADi \cite{andersson2012casadi} and solved with IPOPT \cite{wachter2006implementation}. Computational times range from 3 to \qty{18}{\min} on a commercial laptop (Apple M2 Max, \qty{32}{\giga\byte} RAM). 

%% file: chapters/S4_Results.tex
\section{Results}\label{sec:results}
In this section, we present the optimization results for different case studies. In \cref{sec:case1}, we analyze the interplay between trajectory optimization and wake effects. In \cref{sec:case2}, we compare different game formulations. Finally, in \cref{sec:case3}, we analyze the link between available energy and overtake locations. 
Before diving into the results, we provide some definitions to simplify the notation. The results include case studies involving two agents, $A$ and $B$, whose trajectories are represented in red and blue, respectively. They share the same set of vehicle parameters.
We introduce the notation for gap time as follows: 
\begin{equation}
	t_\mathrm{gap}=t_{\mathrm{gap,rel},B},
\end{equation}
indicating that the gap time is considered relative to $B$. Consequently, if $t_\mathrm{gap}\geq0$ agent $B$ is behind agent $A$, and vice versa. 
The initial temporal position is defined by the initial gap between the two agents. 
\begin{equation}
	t_\mathrm{gap,init}= t_{\mathrm{init},B} - t_{\mathrm{init},A}. 
\end{equation}
Regarding the lateral displacement with respect to the center line, we define it as positive when the agent is on the right side of the center line. The lateral gap is computed with reference to the agent $B$:
\begin{equation}
	y_\mathrm{gap}=y_{\mathrm{gap,rel},B}.
\end{equation}
 As a consequence, when $y_\mathrm{gap}\geq0$ agent $B$ is to the right of agent $A$, and vice versa. 
 Among the case studies we only vary the boundary conditions $t_\mathrm{gap,init}$ and $\Delta{E_{\mathrm{b,target},B}}$.
  
\subsection{Trajectories and interactions}\label{sec:case1}
\begin{figure*}
\centering
\includegraphics[width=2\columnwidth]{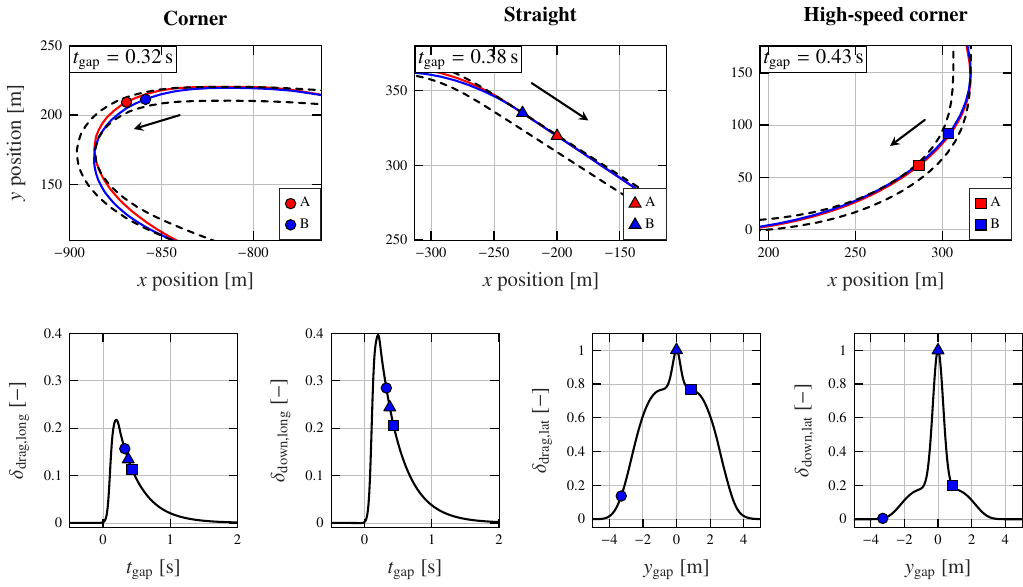}
\caption{Three interaction scenarios on the circuit of Catalunya, Spain. On the top, we show a temporal representation of the agents, their trajectories and track boundaries during a corner (turn 5), a straight (after turn 9) and a high-speed corner (turn 14). The gap times are indicated in each plot. On the bottom, we present the longitudinal and the lateral reductions experienced by agent $B$ for both drag and downforce. The circle, triangle and square in the bottom plots correspond to the scenarios depicted in the top plots.}
\label{fig:CS1}
\end{figure*}

In this case study, we showcase the interplay occurring between the choice of the trajectory and the slipstream effects. Here, we consider the circuit of Catalunya, Spain. In particular, we look at three distinct racing scenarios, where different levels of drag and downforce are required: corners, straights and high-speed corners. \Cref{fig:CS1} shows the resulting trajectories of two agents and the corresponding reduction coefficients in the aforementioned cases. The reduction coefficients are shown only for agent $B$, being behind and affected by the slipstream.  
Although not shown, also the energy management is jointly optimized for both agents. For each scenario, we first outline the expected performance requirements and then analyze the optimization results.

\textit{Corners.} During cornering, the traction force of the car is limited by its maximum grip. To enhance it, modern \gls{f1} cars exploit the suction effect to increase the downforce. Here, the priority is to attain the maximum grip by maximizing the downforce.  

We can observe that $B$ chooses another trajectory than $A$. In longitudinal direction, the drag and downforce reductions are comparable with the other two scenarios. However, the different trajectory results in a lateral gap of $y_\mathrm{gap}=\qty{-3.3}{\meter}$, where $\delta_{\mathrm{down,lat}}=0$. This entirely recovers the downforce, mitigating the effect of dirty air. On the other hand, the drag reduction is very limited, with a total reduction of $C_\mathrm{x,int} = \qty{2}{\percent}$.

\textit{Straights.} In a straight, the traction force is limited by the available power and not by the longitudinal and lateral accelerations. Thus, the cars are not exploiting the downforce. On the downside, the drag power acts against the movement and hinders the acceleration of the car. Since the drag power is proportional to the third power of the velocity, the higher the latter, the more energy is dissipated.
For these reasons, during straights the drag reduction is a desirable feature, whereas a reduced downforce does not come with disadvantages.

The results show that $B$ chooses to remain in the wake of $A$, in order to maximize the drag reduction, with $C_\mathrm{x,int} = \qty{13}{\percent}$. Indeed, the lateral gap is $y_\mathrm{gap} = \qty{0}{\meter}$, and the amount of reduction is solely determined by the longitudinal distance, i.e., the gap time. For the downforce, the same considerations apply, with a total reduction of $C_\mathrm{z,int} = \qty{24}{\percent}$. Despite the massive loss, it does not affect the performance, and the optimal solution prioritizes the drag reduction.

\textit{High-speed corners.} This is a mix of the previous two situations. High downforce is required to achieve a competitive cornering velocity. However, the energy dissipated by the drag is not negligible.

This trade-off is captured by the optimal solution. Indeed, $B$ closely follows $A$, but with a lateral gap of $y_\mathrm{gap}=\qty{0.85}{\meter}$. The total reduction coefficients for this scenario are $C_\mathrm{x,int} = \qty{8.4}{\percent}$ and $C_\mathrm{z,int} = \qty{4.2}{\percent}$ for drag and downforce, respectively. Interestingly, $B$ shifts laterally just enough to limit the downforce reduction while maintaining a relatively high drag reduction.  

Comparing the expected outcomes with the results, we see that the optimal solution closely follows what is intuitively expected from the physical system. By capturing the link between slipstream effects and trajectory optimization, we validate our framework.

\subsection{Comparison of the different games' formulations}\label{sec:case2}
\begin{figure}
\centering
\includegraphics[width=\columnwidth]{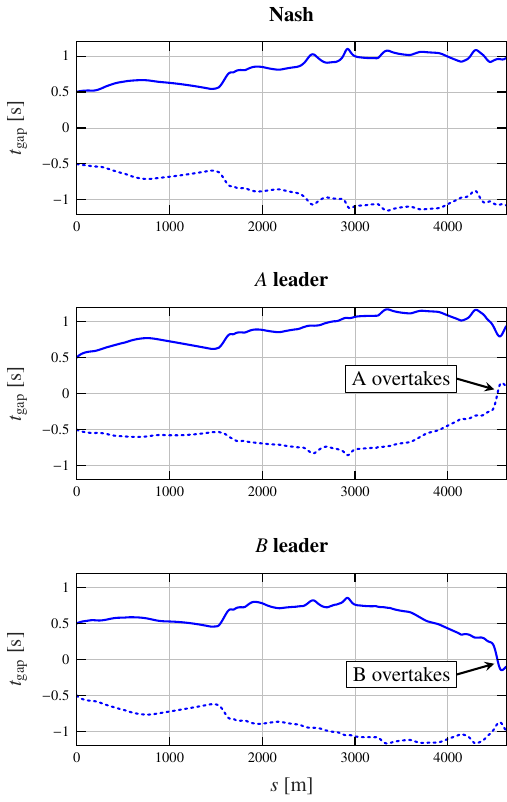}
\caption{Gap time evolution during one lap for 6 scenarios, on the circuit of Catalunya, Spain. The top plot shows two Nash games, once where $B$ starts \qty{0.5}{\second} after $A$ (solid line), and once where $B$ starts \qty{0.5}{\second} ahead of $A$ (dashed line). The second plot shows the same lines for a Stackelberg game where $A$ is the leader, whereas the third plot is a Stackelberg game with $B$ as leader.}
\label{fig:CS2}
\end{figure}

In this section, we compare qualitatively the different games formulations. We consider again the circuit of Catalunya, Spain. \Cref{fig:CS2} presents the solutions for the Nash game, the Stackelberg game with $A$ as leader, and the Stackelberg game with $B$ as leader. Each plot shows two gap time evolutions: When $B$ starts behind by \qty{0.5}{\second} is indicated with the solid line, while when it starts \qty{0.5}{\second} ahead of $A$ is indicated with the dashed line. Note that the leader in the Stackelberg game does not necessarily correspond to the \textit{race} leader.

In the Nash game, changing the starting position delivers the same mirrored solution, up to nonlinearities. Although it might seem a trivial consideration, this confirms the absence of the hierarchical structure in this game formulation. Switching the order of identical agents mirrors the outcome of the game. In both cases, the gap time between the cars increases with an almost identical trajectory. 

The second plot of \Cref{fig:CS2} shows the Stackelberg game solution with $A$ as leader, where the dashed line corresponds to $A$ starting behind, whereas the solid line corresponds to $A$ starting ahead. We observe an asymmetry in the solutions when swapping the position of the agents while keeping their role unchanged, i.e., the leader is always $A$ but it starts behind or ahead. Being the leader empowers the agent to change its strategy, and towards the end even an overtake takes place. 
To find the mirrored solution, we look at the solid line in the third plot of \Cref{fig:CS2}, where $B$ is the leader and starts behind. This is not surprising, since swapping position and role results in the same mathematical problem. However, this helps us to verify our numerical implementation. 

The observed symmetry of the policies validates our game formulations. Permuting agents and/or roles, we obtain the expected solutions. Additionally, the presence or absence of a hierarchical structure affects the decision-making process of the agents. 

\subsection{Battery depletion and overtake location}\label{sec:case3}
\begin{figure*}
\centering
\includegraphics[width=2\columnwidth]{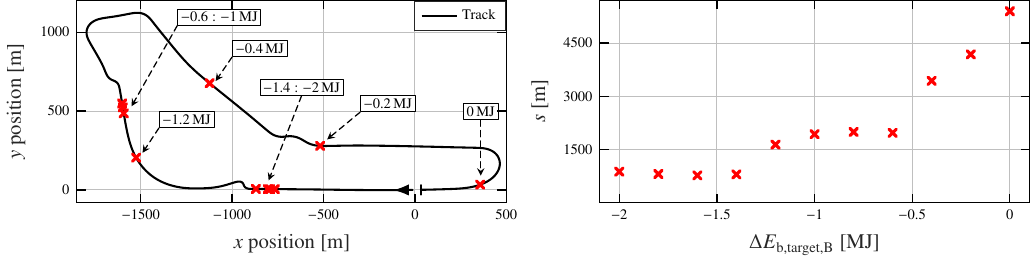}
\caption{Locations of the last overtake, marked with a red cross, for different battery energy targets. In the left plot, we see the Monza circuit with the overtake distribution. The corresponding energy targets are specified in the tags. On the right plot, the distribution of the location is represented directly as a function of the allocated energy $\Delta E_{\mathrm{b,target},B}$. In all the scenarios, the allocated energy by agent $A$ is always $\Delta E_{\mathrm{b,target},A} = \qty{0}{\mega\joule}$. For the sake of clarity, only the last overtake is shown and intermediate ones were neglected.}
\label{fig:CS3}
\end{figure*}

In this section, we analyze the link between available battery energy and overtake locations. To this end, we vary the allocated battery energy for the lap $\Delta E_{\mathrm{b,target},B}$ between $0$ and $\qty{-2}{\mega\joule}$, while keeping $\Delta E_{\mathrm{b,target},A} = \qty{0}{\mega\joule}$. Then, the location of the \textit{final} overtake of $B$ is detected and plotted in \Cref{fig:CS3}. The initial gap time $t_\mathrm{gap,init} = \qty{0.1}{s}$ is the same among all the cases. We consider the circuit of Monza, Italy, where the long straights and high-speed corners enhance the effect of the slipstream interactions.  

The first observation is that the more available energy, the earlier the last overtake. The trend is clearly distinguishable in the right plot of \Cref{fig:CS3}. Starting from $\Delta E_{\mathrm{b,target},B}= \qty{0}{\mega\joule}$, i.e., a charge sustained lap, the last overtake occurs towards the end of the lap at $s = \qty{5400}{m}$. By gradually increasing the available battery energy, i.e., $\Delta E_{\mathrm{b,target},B}$ is more negative, the location of the last overtake shifts towards the beginning of the lap, up to the end of the first straight at around $s = \qty{800}{m}$. 

In terms of drag reduction, one can argue that overtaking too early is not beneficial. This is true to a certain extent, but we have to consider the following: First, the reduction in downforce limits the achievable speed during cornering, resulting in a disadvantage. When ahead, one does not exploit the drag reduction but at least the downforce is completely recovered. Second, the objective is purely a lap time minimization, and waiting behind the other agent despite the additional energy is not lap-time-optimal.  

The overtakes are not linearly distributed along the track according to the available energy, but rather clustered at specific locations. This suggests that some places are better suited to overtaking than others, even with different energy levels. For instance, let us consider the two main clusters for $\Delta E_{\mathrm{b,target},B} = [-0.6, -1] \cup [-1.4, -2] \unit{\mega\joule}$. For those points, the overtakes occur towards the end of a straight or a high-speed corner. The drag reduction experienced along that section is exploited to gain a velocity advantage. Towards the end of the straight, the agent moves to the side and overtakes. This is a typical maneuver in \gls{f1}, usually undertaken in combination with the \gls{drs} to enhance the drag reduction. Additionally, overtaking at the end of the straight is strategically advantageous, because the overtaken car ends up in the wake of the leading car. This results in a reduction of downforce and grip, making it less likely for the overtaken car to re-overtake. Eventually, it remains important to distinguish the single clusters, determined by the strategical exploitation of the wake effect, from their distribution, still influenced by the available energy.

Overtakes for the cases $\Delta E_{\mathrm{b,target},B} = \{0, -0.2\} \unit{\mega\joule}$ are more related to energy management. They happen at the beginning of a straight, indicating that the extra energy, allocated or saved due to the drag reduction, is exploited to extend the \gls{mguk} boosting time with respect to the competitor (not shown here). It is interesting to notice that even when the two agents have the same battery energy allocation, i.e., where $\Delta E_{\mathrm{b,target},B} = {0} \unit{\mega\joule}$, an overtake still occurs. The energy saved thanks to the drag reduction over the entire lap is used to effectively overtake the other agent. This underlines the influence of the wake interaction on the energy management.

This analysis further validates the framework in its completeness, by linking the energy management to multi-agent interactions. Furthermore, it is possible to distinguish between the effects of trajectory optimization, slipstream interactions or battery energy target.

%% file: chapters/S5_Conclusions.tex
\section{Conclusion and Outlook}\label{sec:conclusion}

In this paper, we presented a complete framework to describe and solve multi-agent interactions in the context of motorsport racing. By significantly extending the contributions of \cite{fieni2024game}, our current work advances this model by fully describing the wake effect on trailing cars. Specifically, the model now incorporates both drag and downforce reductions as functions of the longitudinal and lateral proximity to the leading car. Key enhancements include a dynamic trajectory model that enables adaptive paths and energy management strategies to either exploit or mitigate aerodynamic effects. Additionally, the inclusion of collision avoidance constraints allows for a more accurate replication of realistic multi-agent behavior in dynamical environments. 

By means of three case studies, we isolated and highlighted the interplays occurring in this complex environment. The first analysis across typical \gls{f1} scenarios—corners, straights, and high-speed corners—revealed distinct strategies: Trailing vehicles prioritize lateral displacement in corners to maximize grip, exploit maximal drag reduction on straights, and balance both effects in high-speed corners. Afterwards, we commented on the expanded game-theoretic approach. While the previous paper \cite{fieni2024game} only considered a fixed leader-follower Stackelberg game, our current study enables to change the roles or to formulate a Nash game. The hierarchical approach enables non-symmetrical results, with the leader acting as the primary decision-maker achieving superior performance by anticipating the follower's responses. Conversely, Nash solutions produced symmetric outcomes, validated by mirrored results when the agents' roles are reverted. Given the similar formulations, we can compare the outcome of the games. We also introduced a method to find a better local optimum of the \gls{nlp} by exploiting a mathematical property of game theory. The last analysis studied the influence of the energy management strategy on overtake locations. Apart from the expected trend following the available battery energy, we could also distinguish the joint influence of trajectory optimization and slipstream effects.

Future research could extend this framework to multi-lap strategies across an entire race, including strategic studies to incorporate tire-saving models in the optimization. The problem could be extended to team-based cooperation, for instance, during qualifying sessions when slipstreams might benefit the trailing car, or in race scenarios where team collaboration could facilitate overtaking other cars. Another promising direction involves the inclusion of stochastic models to compute mixed strategies. This would enable the study of robust policies capable of accounting for uncertainties in a competitor's behavior.

Beyond motorsport, this work has potential applications in domains such as autonomous driving and robotics, where the presence of other interacting entities affects the decision-making process or the single-agent optimal policy. Game-theoretic-based controller could be developed, which are known to outperform classical \gls{mpc} solutions. In this direction, decentralized decision-making simulations can be performed. Agents could be free to choose, negotiate or adapt their role during the game. Different combinations of roles can also be tested, for example a leader playing against another leader \cite{cinar2024does}. Multi-layered systems' optimization can also benefit from a Stackelberg formulation, where lower levels comply with higher levels while simultaneously optimizing their own objective.

%% file: chapters/S6_Appendix_1.tex
\section{Single-leader-multi-follower Stackelberg Game}\label{sec:appendix_1}
Although not directly relevant in this study, it is possible to combine the Nash and Stackelberg games reformulations. Their concept can be extended to single-leader-multi-follower scenarios, with followers not subordinate to each other. \Cref{prob:multiFollower} presents the mathematical formulation.

\begin{problem}\label{prob:multiFollower}
The \gls{kkt}-based reformulation of the single-leader-multi-follower game reads
\begin{alignat*}{2}
 & \min_{\mathbf{x}_{L}, \mathbf{u}_{L}, \mathbf{x}_{F}, \mathbf{u}_{F}} J_{L}(\mathbf{x}_{L},\mathbf{u}_{L}, \mathbf{x}_{F}) + \sum_{i = 1}^{n} J_{F,i}(\mathbf{x}_{F,i},\mathbf{u}_{F,i}, \mathbf{x}_{-i})\\
 & \text{subject to:} \\
		& \quad \quad \mathbf{g}_{L}(\mathbf{x}_{L},\mathbf{u}_{L}, \mathbf{x}_{F}) \le 0, \\
		& \quad \quad \mathbf{h}_{L}(\mathbf{x}_{L},\mathbf{u}_{L}, \mathbf{x}_{F}) = 0,\\
		& \quad \quad \mathcal{KKT}_{F,1}(\mathbf{x}_{F,1},\mathbf{u}_{F,1}, \mathbf{x}_{-1},\boldsymbol{\lambda}_{F,1},\boldsymbol{\mu}_{F,1}),\\
		&\quad \quad \vdots \\
		& \quad \quad \mathcal{KKT}_{F,n}(\mathbf{x}_{F,n},\mathbf{u}_{F,n}, \mathbf{x}_{-n},\boldsymbol{\lambda}_{F,n},\boldsymbol{\mu}_{F,n}),
\end{alignat*}
which is a single-level nonlinear program.
\end{problem}

\section{Equilibria properties}\label{app:equilibria}
When comparing the achieved cost of leader and follower with their Nash solution, we can distinguish between three different cases, each one with its own (dis-)equilibrium \cite{simaan1977equilibrium}. We point out that disequilibria are also feasible solutions.
\begin{description}
\item[Case 1.] None of the Stackelberg solutions is dominant. The leader always incurs a better cost than the Nash solution, but it is not the case for the follower. The latter has no incentive to play the follower, and thus each Stackelberg solution is a disequilibrium. 
\item[Case 2.] Only one Stackelberg solution is dominant. Both leader and follower incur a better cost than the Nash solution, and both agree that this Stackelberg solution is an equilibrium.
\item[Case 3.] Both Stackelberg solutions are dominant. Hence, for both there is no need to play Nash, since they can in any case achieve a better cost. Here, we need to look at the \textit{relative} Stackelberg values and distinguish again in three cases \cite{simaan1977equilibrium,basar1973relative}. 
\begin{description}
\item[Concurrent solution.] The player $i$ incurs a better cost as a follower than as a leader. The leadership of player $-i$ is thus better, and both agents agree on that solution, defining the solution as a Stackelberg equilibrium. 
\item[Nonconcurrent solution.] Both players perform better under their own leadership. Since no agreement is found, these solutions are disequilibria.
\item[Stalemate solution.] Both players have a better cost when they are followers. An agreement cannot be found, and both solutions are disequilibria.
\end{description}
\end{description}
Although not directly useful, understanding these concepts allows for a deeper understanding of the logic behind each game. Studying the (dis-)equilibria can lead to conclusions on how an agent should behave in certain situations to achieve a better result. Depending on the game properties, such as the setup or the objective, we can characterize the agents' natural behavior for that particular sport.